\documentclass{aa}

\usepackage{natbib,twoopt}
\usepackage{graphicx}
\usepackage{wrapfig}
\usepackage{float}
\usepackage{txfonts}
\usepackage{multirow}
\usepackage{tablefootnote}
\usepackage{amsmath}
\usepackage{siunitx}
\usepackage{indentfirst}
\usepackage{subfig}
\usepackage{blindtext}

\usepackage[dvipsnames]{xcolor}
\usepackage{xcolor}
\definecolor{darkgoldenrod}{rgb}{0.72, 0.53, 0.04}

\usepackage[colorlinks = true,
            linkcolor = RoyalBlue,
            urlcolor = darkgoldenrod,
            citecolor = RoyalBlue,
            anchorcolor = RoyalBlue, 
            draft = false]{hyperref}

\newcommand{\comment}[1]{}

\begin{document} 
   \title{Discovery of the Polar Ring Galaxies \\ with deep learning}

  \author{Dobrycheva D.V.$^{1}$, Hetmantsev O.O.$^{1,2}$, Vavilova I.B.$^{1}$, Shportko A.$^{3}$, Gugnin O.$^{2}$, Kompaniiets O.V.$^{1}$}

   \institute{
$^{1}$Main Astronomical Observatory of the NAS of Ukraine, 27, Akademik Zabolotnyi St., Kyiv, 03143, Ukraine\\
$^{2}$Taras Shevchenko National University of Kyiv, Hlushkov Ave., 4, Kyiv, 03127, Ukraine \\
$^{3}$Northwestern University, 633, Clark St., Evanston, IL 60208, Chicago, USA \\
}

\titlerunning{Discovery of the Polar Ring Galaxies}
\authorrunning{Dobrycheva D.V., et al.}

   \date{Received April 5, 2025; accepted }
  \abstract
   {Polar ring galaxies (PRGs) play an important role in understanding the evolution of galaxies, especially as unique cases of gas accretion and merging process between early and late morphological galaxy types. Regardless of their spectacular shape, these objects are very few in number and hard to find. Most of them were visually discovered, and then several of them were photometrically validated and kinematically confirmed.
   }
   {The aim of our research is to create a catalog of strong and good candidates for PRGs using existing catalogs of PRGs; to develop an image-based approach with machine learning methods for the search and discovery of PRGs in a big sky survey; to explore the capability of the CIGALE software for determining their multiwavelength properties.   
   }
  {For the first time, we applied a deep learning method to the search for PRGs. We visually inspected galaxies from existing catalogs of PRGs to create a training sample based on high-quality SDSS images. Since the resulting training sample was extremely small (87 strong and good PRGs), we applied augmentation, image segmentation, and ensemble learning techniques. However, most effective method was transfer learning with its ability to enlarge the training sample by synthetic images generated by GALFIT. To examine deep learning approach for finding new PRGs we used the SDSS catalog of galaxies at $z$ < 0.1. The method with synthetic images showed that even with overtraining we were able to find galaxies with a ring pattern.  
  }
   {Our deep learning approach has resulted in the discovery of three PRGs (SDSS J140644.42+471602.0; SDSS J133650.48+492745.3; SDSS J095717.30+364953.5).   
   Also, we visually inspected the Catalog of the SDSS Ring galaxies at $z$ < 0.1 (it's a part of a catalog by \cite{Vavilova2023}) and discovered four PRGs among $\sim$2,200 ring galaxies (SDSS J095851.32+320422.9; SDSS J104211.05+234448.2; SDSS J162212.63+272032.2; SDSS J104600.10+090627.2). 
   One of the discovered galaxies with transfer learning, SDSS J140644.42+471602.0, was studied with CIGALE software to determine its spectral energy distribution in IR - UV bands. The current SFR is 71 $M_{\odot}$ per year, although the lack of FUV data limits this estimate. The total stellar mass is 8.34$\times 10^{10}$ $M_{\odot}$. The predominance of an old stellar population (two-thirds of the total mass) suggests that this PRG is undergoing interaction process.
   Finally, we present a catalog of 179 visually inspected PRGs. Being supplemented with new discovered objects, it will become quite useful as for CNN approach and theoretical studies. Our strategies represent valuable opportunities for future development, aiming to push the boundaries of our current deep learning model and achieve more reliable PRG identification in big sky surveys.   
   }
{}
   \maketitle

\section{Introduction}
\noindent
There are a few dominant theories of the formation of galaxies with polar rings (PRGs). 

The first theory suggests that two galaxies formed this peculiar system due to the merger process, where the central and dominant galaxy is an early-type galaxy, and the ring around it is a late-type galaxy \citep{Schweizer1983, Bekki1997, Bekki1998}. For example, recently, \cite{Akhil2024b} explored the globular cluster system of PRG, NGC 4262, and found evidence of previous interactions within the host galaxy, supporting that PRGs are an intermediate phase of merging with quiescent galaxies. The second theory states that the polar ring is formed by the gas accretion of the donor galaxy onto the dominant galaxy. \cite{Bournaud2003} studied these two theories using N-body simulations, including gas dynamics and star formation processes. They concluded that the accretion scenario is the most likely. \cite{Ordenes2016} explored the tidal interaction between NGC 3808 A and B, proposing it as a PRG mechanism. Both theories need to explain the problem of the central disc retaining its rotational support in systems like NGC 4650A, where polar discs are significantly massive.

The interesting result was published by \cite{Haud1988} that our Milky Way galaxy also has probably a faint polar ring related to the Large Magellanic Cloud. Modern studies did not confirm this hypothesis (see, for example, \cite{Xu2023}), as well as the Gaia discovered that the last major merger of the Milky Way with Gaia Sausage Enceladus was 8-11 billion years ago \citep{Helmi2018, Naidu2021, Vavilova2024}.

\begin{figure*}[t]
	\centering
	\includegraphics[width = 0.25\linewidth]{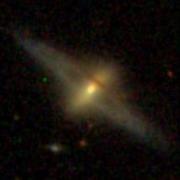}
    \includegraphics[width = 0.25\linewidth]{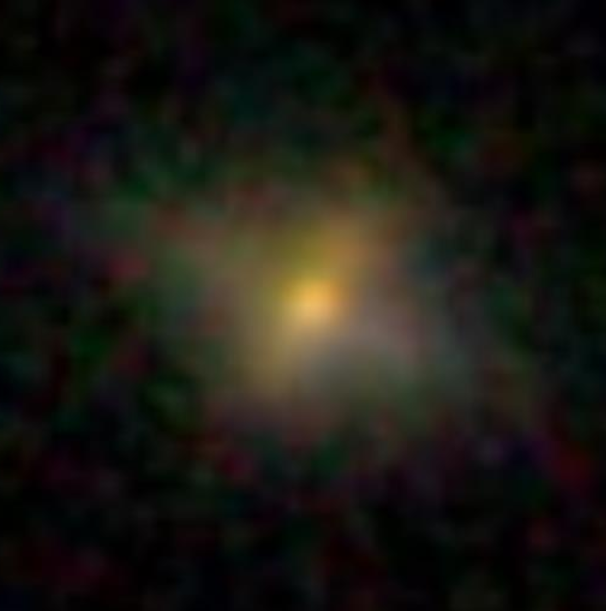}
    \includegraphics[width = 0.25\linewidth]{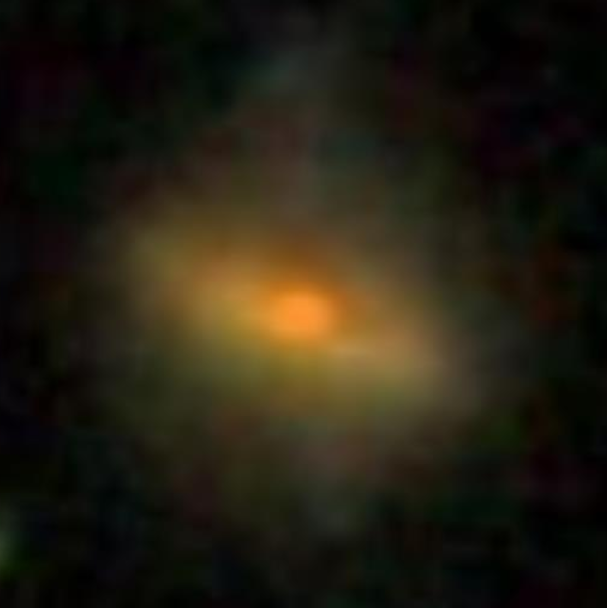}
	\caption{Examples of PRG candidates identified by visual inspection: (a) Strong, RA: 24.7301, DEC: -7.7654, z=0.035, kinematically confirmed by \cite{Whitmore1990}; (b) Good, RA: 10.2643, DEC: -9.9411, z=0.037 \citep{Moiseev2011}; \\ (c) Weak, RA: 161.5985, DEC: 6.6194, z=0.028 \citep{Moiseev2011}.}
	\label{PRG_SGW}
\end{figure*}

\begin{figure*}[t]
	\centering
    \includegraphics[width = 0.20\linewidth]{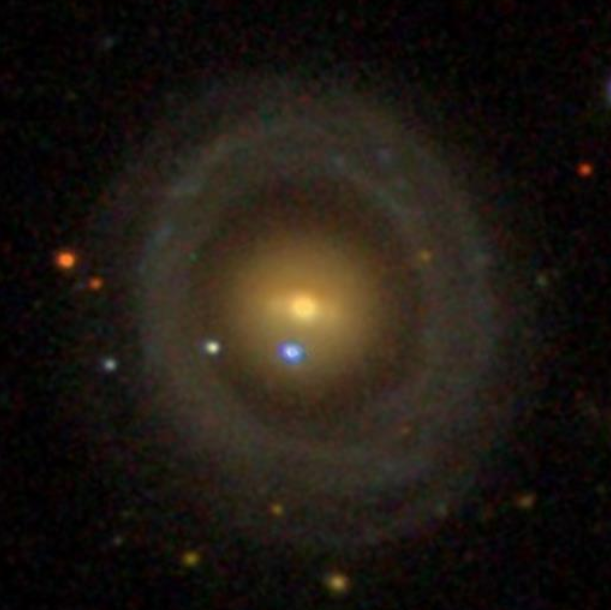}
	\includegraphics[width = 0.20\linewidth]{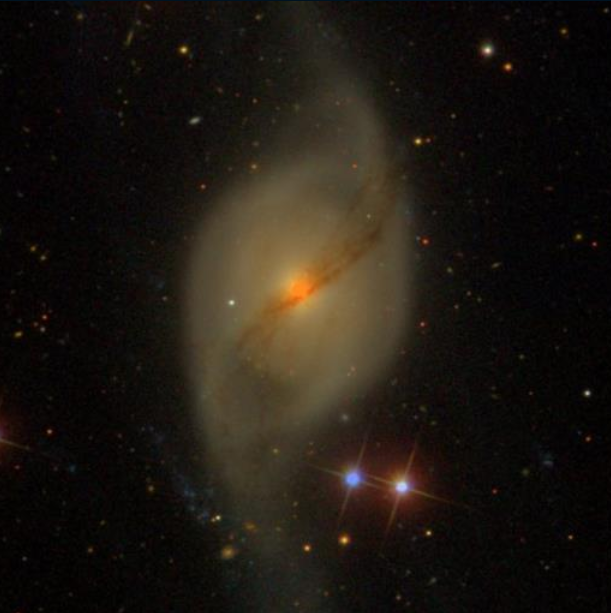}
    \includegraphics[width = 0.20\linewidth]{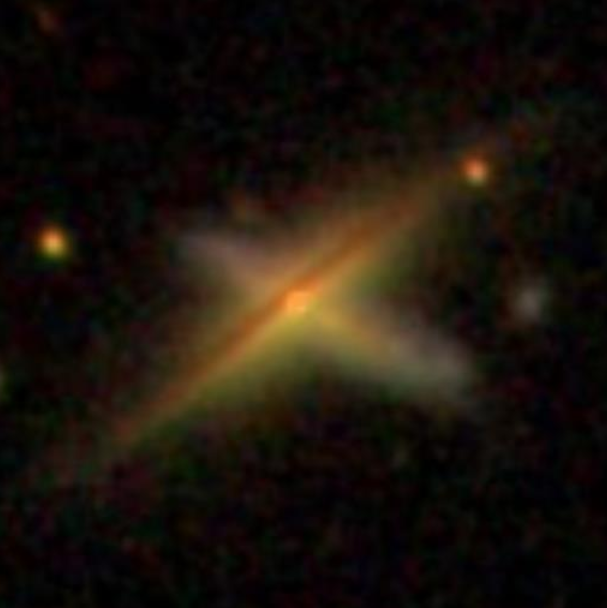}
    \includegraphics[width = 0.20\linewidth]{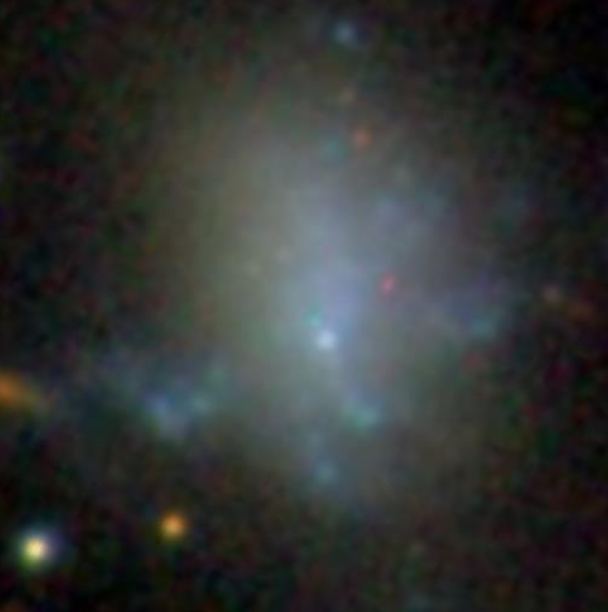}
	\caption{Non-PRGs identified by visual inspection: (a) Ring, RA: 240.3706, DEC: 19.3598, z=0.015 \citep{Whitmore1990}; (b) Dust lane, RA: 173.1451, DEC: 53.0679, z=0.0033 \citep{Whitmore1990} more detail in \cite{Akhil2025}; (c) Line of sight, RA: 349.5518, DEC: 4.1851, z=0.040 \citep{Whitmore1990}; (d) Irr, RA: 125.9668, DEC: 14.7521, z=0.007 \citep{Whitmore1990}.}
	\label{PRG_DL_R_LS}
\end{figure*}

The third theory suggests that cold gas accretion along cosmological filaments may form and continue to feed polar discs \citep{Maccio2006}. \cite{Brook2008} used high-resolution cosmological simulations of PRG formation. According to their model, PRGs form through the continuous gas accretion, whose angular momentum is misaligned with the central galaxy. Additionally, \cite{Snaith2012} studied the dark matter halo shape in PRGs, and showed that the inner regions of the halo are often flattened perpendicular to the old stellar disc, aligning with the polar disc out until the virial radius. They also claim that the polar disc structure is the result of the direction of gas infall from the filament, with continuous evolution of the angular momentum of the infalling gas and stars. Also, \cite{Maccio2006} showed that the formation of such a system can occur naturally in a hierarchical scenario, where most low-mass galaxies are assembled through the accretion of cold gas infalling along megaparsec-scale filamentary structures. \cite{Stanonik2009} have found an isolated PRG between two voids, which demonstrates slow accretion of the HI material: the central stellar disc appears relatively blue, with faint near-UV emission, and is oriented roughly parallel to the surrounding wall, implying gas accretion from the voids. This result favors the isolation criteria choice, considering the morphology, multi-wavelength properties, and kinematics/dynamics of PRGs among isolated and filamentary configurations of void galaxies \citep{Vavilova2009, Beygu2011, Melnyk2012, Freitas2014, Pulatova2015, Dobrycheva2018, Mahdi2021}. For example, \cite{Finkelman2012} found that PRGs preferentially settle in low-density environments compared to other early-type galaxy samples. They explained this result by a natural selection effect; in many PRGs, the luminous polar structure extends much farther than the main body and could be destroyed more easily by close-neighbor galaxies. 

The theories mentioned above and the research that was conducted have suggested possible mechanisms for the formation of PRGs. The diversity of results only confirms that the individual characteristics and group trends of these spectacular and rare objects are not yet fully understood. The existing PRG catalogs and lists are compiled primarily by visual inspection and validation \citep{Whitmore1990, Moiseev2011, Reshetnikov2019}. The photometrical studies are often difficult because these galaxies have a complex structure of two morphological types; the polar ring may be dimmed or destroyed by tidal effects; in the case of nearby PRG, the telescope aperture size covers part of the galaxy or the overlapped central part, etc. However, \cite{Lackey2023, Lackey2024} have demonstrated that in several cases of PRGs, a multiwavelength analysis of polar rings can be exploited to estimate sizes, ages, masses, and star formation rates (SFRs) using photometry and spectral energy distributions (SEDs) in the optical and UV bands. 

The aim of our paper is as follows: to create a catalog of strong and good candidates for PRGs using available lists of PRGs; to develop an image-based approach with machine learning methods for the search and discovery of PRGs in a big sky survey; to explore the capability of the CIGALE software for determining the multiwavelength PRGs' properties. 

We describe a new catalog of PRGs compiled after visual inspection in Section 2; the CNN approach for PRG identification is given in Section 3; the procedure with transfer learning through simulated images and results for the discovered PRG candidates is presented in Section 4; visually discovered PRGs candidates are briefly noted in Section 5; the multiwavelength analysis of one of the strongest PRG candidates found by us, SDSS J140644.42+471602.0, is exploited as an example with CIGALE software in Section 6; the conclusions are summarized in Section 7. 

\section{Catalog of Visually Inspected PRGs}
\noindent
\subsection{Available observational data for the formation of PRG sample}

Our preliminary sample of 463 candidates to PRGs was compiled from existing polar ring galaxy datasets and covered four sources: Atlas of Polar Ring Galaxies by \cite{Whitmore1990}; a New Catalog of Polar Ring Galaxies by \cite{Moiseev2011}; a list of New candidates to polar ring galaxies by \cite{Reshetnikov2019}; results by \cite{Skryabina2024}.

The Atlas of Polar Ring Galaxies by \cite{Whitmore1990} was compiled by visually inspecting galaxies from the Uppsala General Catalog of Galaxies (\cite{Nilson1973}), the ESO Catalog of Galaxies (\cite{Lauberts1982}), and the Southern Atlas of Peculiar Galaxies (\cite{Arp1987}), which were commented on as having an unusual needle-like or similar shape. In total, this Atlas included 157 objects divided into four categories: kinematically confirmed PRGs (6 galaxies), good candidates (27 galaxies), possible candidates (73 galaxies), and possibly related systems (51 galaxies). These authors also noted that about 5 \% of S0 galaxies have, or have had, a polar ring. 

A new catalog of PRGs by \cite{Moiseev2011} includes 275 objects. They have visually examined around 42,000 Galaxy Zoo images with various probabilities to belong to a specific type and searched the Internet forum dedicated to ring galaxies. These authors divided the Sloan-based Polar Ring Galaxy Catalog into four types: best candidates (70 galaxies), good candidates (115 galaxies), related objects (53 galaxies), and possible face-on rings (37 galaxies). 

\cite{Reshetnikov2019} prepared a list of 31 new candidates for PRGs by visual inspection and analyzing the Galaxy Zoo Project discussion boards.

With the launch of new sky surveys, the units of PRGs were found. \cite{Skryabina2024} examined deep optical images of edge-on galaxies from SDSS Stripe 82, DESI Legacy Imaging Surveys, and Hyper Suprime-Cam Subaru Strategic Program images, and have yet to find three galaxies that demonstrate dim polar rings. We added these galaxies, SDSS RA: 333.748, DEC: 1.020; SDSS RA: 326.580, DEC: -0.208;  SDSS RA: 29.743, DEC: -0.490, to our preliminary sample. \cite{Akhil2024} discovered the PRG DES J024008.08-551047.5 (DJ0240) with the Dark Energy Camera Legacy Survey. 


Also, \cite{Bahr2024} announced the selection of 102 galaxies with polar structures (polar rings, halos, bulges, and forming polar structures), reviewing over 18,000 galaxies in the same sky surveys (lists are not yet published). 

\subsection{Visual Inspection of PRG sample}

We visually inspected 463 objects from our PRG sample in pursuit of multiple goals: to select candidates for our Catalog of Inspected PRGs; to determine strong and good quality PRGs' images that would be most suitable for training the CNN; to identify new PRGs within ``The image-based morphological catalogs of SDSS galaxies at 0.02<z<0.1'' by \cite{Vavilova2021}. 

Consequently, a visual inspection was performed for 357 objects, which matched good quality SDSS images \citep{Blanton2017, Vavilova2020}. In the visual inspection process, we split this sample into the following categories: 

\textbf{Strong candidates:} 26 galaxies with polar rings, which are strictly visible. However, we know from the original catalogs that most are not yet kinematically confirmed as PRGs. Examples of strong PRGs are shown in \autoref{PRG_SGW} (a). 

\textbf{Good candidates:} 61 galaxies with polar rings, which are not as strictly visible as in strong PRGs because of their apparent size. Examples of good PRGs are shown in \autoref{PRG_SGW} (b).

\textbf{Weak candidates:} 80 galaxies, in which the polar rings are hardly differentiated but are still visible. Examples of weak PRG candidates are shown in \autoref{PRG_SGW} (c).

\textbf{Ring galaxies:} 32 galaxies with rings that are not oriented in the polar plane. Examples of ring galaxies found in our catalog of PRGs are shown in \autoref{PRG_DL_R_LS} (a). \textbf{Merging galaxies:} 20 galaxies experiencing collisions with each other. \textbf{Dust lane galaxies:} 31 galaxies exhibiting a dust lane, which could be confused with having a polar ring (\autoref{PRG_DL_R_LS} (b)). \textbf{Line-of-sight galaxies:} 11 instances, where the closer galaxy visually covers the farther one, making them appear similar to PRGs (\autoref{PRG_DL_R_LS} (c)). \textbf{Irregular galaxies:} 39 irregular galaxies with no polar ring were found in our sample (\autoref{PRG_DL_R_LS} (d)). \textbf{Non-ring galaxies: }54 galaxies that do not exhibit any ring structure. \textbf{Quasars:} 1 quasar PG 1100+772 (3C 249.1) was found in the Atlas by \cite{Whitmore1990}, which was previously suggested by \cite{Stockton1983} as PRG. 

\subsection{Training sample for machine learning image-based approach}

Our training sample consisted of 87 PRGs, which were classified as strong and good objects with high-quality SDSS images. Of course, it is an extremely small sample for training a machine learning model. Such a limited sample size could lead to overfitting of machine learning methods and poor generalization. Therefore, we must apply methods that expand the training sample while keeping the key features of the images. In the following sections, we present and discuss different approaches designed to increase the number of training examples and improve the model performance.

\begin{figure*}[h]
    \centering    \includegraphics[width=\textwidth]{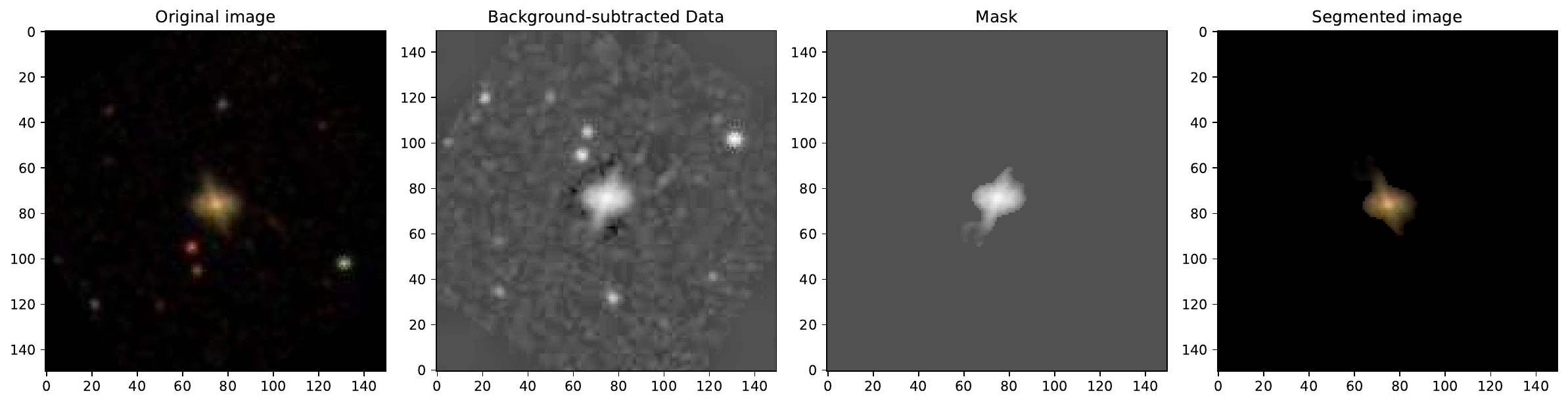}
    \caption{Illustration of the segmentation process applied to a sample image from the training dataset. The leftmost image displays the original data point. The subsequent image demonstrates the background-subtracted data. This is followed by the mask delineating the defined central source. The rightmost image represents the segmented output, which will be used as input for the neural network. The procedure was executed within a single color channel.}
    \label{fig:segm_showcase}
\end{figure*}

\section{CNN image-based approach for PRG Identification} 
\noindent
\subsection{Data generation using augmentations} 

To solve the problem with the imbalance of the training sample, we utilized torchvision transforms \cite{Torchvision2016} to generate augmented data using a composed set of random horizontal and vertical flips with random rotations. This approach was empirically selected after testing different augmentation combinations such as Color Jitters, Random Perspectives, and other methods. We performed these augmentations to increase the dataset size and improve the model's generalization ability from the limited data available. 

Both original and generated images were resized to 80x80 pixels using bicubic interpolation \citep{Keys1981}. We have chosen this resolution as the minimal shape of the input data while ensuring feature preservation. Bicubic interpolation was preferred for downscaling because it maintains the original information better than bilinear or nearest neighbor, which can introduce artifacts and degrade image quality.

\subsection{Image Segmentation} 

The positions of each of the 87 PRGs are located in the central regions of the images. The other information around them is not necessary for the mentioned task. So, to focus only on the relevant features and erase the unnecessary data, we defined a custom transformation to segment the central object using the Astropy \citep{Price2018} and Photutils \citep{Bradley2024} Python packages.

The main routine for this transformation involved several steps. Firstly, we subtracted the median background to increase the contrast between the galaxy and the surroundings. Then we convolved the image with a Gaussian kernel to smooth it and reduce the noise. Following this, we applied source detection to identify objects within the image that exceeded a defined threshold. This threshold-based detection helped to pinpoint the central object accurately.

After detecting the sources, we normalized the images to ensure uniformity across the dataset. Finally, we extracted the central object, placing it against a black background. So, for the input for our model, we have used this processed image, now containing only the central galaxy (as illustrated in \autoref{fig:segm_showcase}).

\subsection{Architecture and training} 

The architecture of our main model consists of two successive convolutional blocks with ReLU activation functions, dropout layers, and one pooling layer. Following the convolutional block, the classifier section consists of a flattening layer, a dense layer, and a sigmoid activation function.

The training process utilizes a Binary Cross Entropy Loss function and the AdamW optimizer \citep{Loshchilov2019}, complemented by a learning rate scheduler that automatically reduces the learning rate upon encountering a plateau. Through an extensive grid search, we optimized the hyperparameters to include one double convolutional block, 10 hidden units, an initial learning rate of $10^{-4}$, weight decay for L2 regularization of $10^{-3}$, and a dropout rate of 0.5. So, the network converges within 100 epochs, achieving an accuracy of up to 90\,\%.

\begin{table}[h!]
\centering
\caption{Resulting confusion matrix for the test dataset. As one notices, it reveals a predominance of True Positive and True Negative values, with fewer instances of false classifications.}
\begin{tabular}{|c|c|c|}
\hline
\textbf{}                  & \textbf{Others} & \textbf{PRG} \\ \hline
\textbf{Others} & 86                           & 7                          \\ \hline
\textbf{PRG}    & 15                           & 75                         \\ \hline
\end{tabular}
\label{conf_mat1}
\end{table}

The model's performance is illustrated in \autoref{conf_mat1}, which presents a confusion matrix derived from the additionally generated images in the test set. The model achieves an accuracy of $\approx$88\%. Despite demonstrating excellent performance on the training, validation, and test sets, the model struggles to generalize across the SDSS catalog by  \citep{Vavilova2021}, where we aimed to identify PRGs. The model classified 48,974 galaxies as PRGs out of the entire sample, which seems implausibly high.

\subsection{Model ensemble and potential improvements} 

To further explore the capabilities of our approach, we conducted two additional independent training sessions using slightly different architectures. The first was a simpler model without convolutional layers, while the second retained the same architecture as the main model but included the segmentation step. By combining the predictions from these models into an ensemble with equal weighting, we aimed to improve classification performance. As a result of this ensemble approach, 494 galaxies were classified as PRGs with scores the higher than 0.9. However, upon visual inspection, none of these galaxies exhibited the characteristics of PRGs. The programming code is available by GitHub\footnote{\url{https://github.com/alexgugnin/prg_detection}}.

Another independent approach was to train a shallow neural network with one hidden layer. To address the problem of an insufficient dataset, the number of PRG images was increased by 25 times through data augmentation. The model made many false-positive predictions when applied to a representative SDSS dataset. Through visual inspection, 25 potential candidates with an accuracy of 0.94 to be PRGs or related objects were selected. Most of these galaxies do not fit the PRG classification since they are spirals with untwisted arms. 

\begin{figure}[h!]
    \centering
    {{\includegraphics[width=6cm]{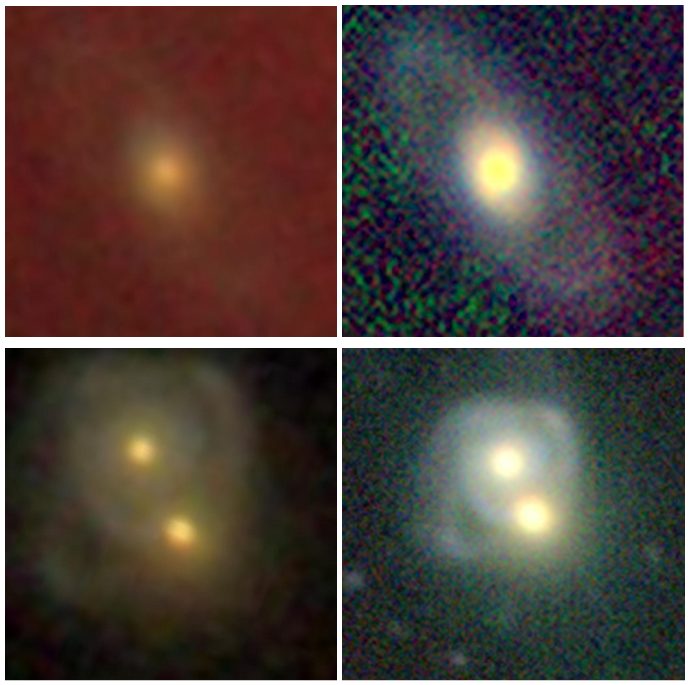} }}   
    \caption{Discovered PRGs by machine learning with accuracy 0.94. 
    First row: (left) SDSS image and (right) DESI Legacy Survey image of PRG with RA: 204.2103, DEC: 49.4625, z=0.097. 
    Second row: (left) SDSS image and (right) DESI Legacy Survey image of double PRG system with RA: 149.3220, DEC: 36.8315, z=0.053 (center) and RA: 149.3231, DEC: 36.8332, z=0.054 (top).}
    \label{PRG_A}
\end{figure}

We considered one more galaxy as a PRG candidate (SDSS, RA: 204.2103, DEC: 49.4625, z=0.097). Despite its SDSS image overlapping with an artifact, a deep learning approach allowed us to discover this PRG (see \autoref{PRG_A}, first row, where SDSS (left) and DESI Legacy Survey (right) images are given). 

We also focus on two images of galaxies, which our deep learning approach recognized as the PRGs. One is the PRG from the catalog by \cite{Moiseev2011} assigned by them as a good candidate. Another is a pair of galaxies (second row in \autoref{PRG_A}). This pair has a ring pattern, one galaxy is elliptical (SDSS RA: 149.3220, DEC: 36.8315, z=0.053), and the second galaxy is a late-type galaxy (SDSS RA: 149.3231, DEC: 36.8332, z=0.054). Since they are at a reasonably close distance, we suggest that the beginning of the interaction is fixed, and we found an example of a PRG candidate in support of the merging mechanism (see \cite{Akhil2024b}). That is why this pair of objects should be investigated in more detail.

The programming code of the CNN model, which allowed the discovery of two PRGs, is available on GitHub\footnote{\url{https://github.com/Antebe/Polar-ring-galaxies}}.

\section{Transfer learning via simulated images}
\noindent
The biggest challenge in classifying PRGs is training a neural network on a severely insufficient dataset. To combat this issue, we decided to use transfer learning. Given that a pre-trained model exists, it allowed us to train a neural network on a smaller dataset. 

\subsection{Simulated galaxies with GALFIT}

GALFIT (\cite{Peng2002}) is mainly used to fit light profiles to galaxies. However, it can also generate such light profiles by manually setting parameters via input files, allowing us to create a dataset of simulated PRG images. 

\cite{Ghosh2020} developed the GalSim code, which generates input files and then runs GALFIT to automatically generate the required number of images with randomized parameters. Later, \cite{Krishnakumar2022} used the modified GalSim code to generate a sample of simulated galaxies with rings. The process of simulating images of galaxies with GalSim goes as follows: first, we create a chosen number of GALFIT input files with simulation parameters, then GALFIT runs on these files, producing the simulated images, and, lastly, a noise is added to the images, and conversion from FITS to JPG is done.

To simulate the central galaxy, we used the Sérsic function, which is widely used in galaxy morphology. The Sérsic profile is defined as \citep{Peng2002}: 
$$ \Sigma(r) = \Sigma_e \exp \left[ -\kappa \Biggl( 
\left( \frac{r}{r_e}\right)^{1/n} - 1\Biggr) \right]$$,
where $\Sigma_e$ is the pixel surface brightness at the radius $r_e$, representing the radius at which half of the galaxy's flux is contained.$n$ is the Sérsic index of the galaxy, which controls where the light of the galaxy is concentrated, and $\kappa$ depends on this parameter to ensure that half of the flux stays within radius $r_e$ from the center. For accurate simulation, the polar-ring component must be truncated in the center. This is achieved by multiplying the Sérsic function of the polar ring component by a hyperbolic tangent truncation function in GALFIT. 

We have taken the parameters from \cite{Krishnakumar2022} for generating non-ring and ring galaxies. We modified them to simulate PRGs by rotating the position angle of the ring component by $90^{0}$ and the axis ratio of the ring component in such a way that the image resembles more closely a PRG (parameters are shown in \autoref{tab:galfit-parameters}). In total, 1,000 PRG images and 3,000 non-PRG images were simulated using GALFIT, which is enough to train a neural network. The model was trained on the synthetic dataset over eight epochs utilizing the Adam optimizer (\cite{Kingma2015}) with the learning rate of $10^{-4}$, reaching convergence on training and validation sets. 

\begin{table}[h!]
\centering
\caption{GALFIT parameters for the inner component and polar ring component of the PRG}
\begin{tabular}{|c|c|c|}
\hline
\textbf{Parameter}      & \textbf{Inner Galaxy} & \textbf{Polar Ring} \\ \hline
Sérsic index            & 3.0                   & 1.7                 \\ \hline
Half-light radius (kpc) & 8.0 — 12.0            & 2.0 — 6.0           \\ \hline
Axis ratio              & 0.5 — 0.75            & 0.3 — 0.7           \\ \hline
Position angle (pix)    & 0.0                   & 90.0                \\ \hline
\end{tabular}
\label{tab:galfit-parameters}
\end{table}

\subsection{Transfer learning and final search}

Transfer learning includes the "freezing" of most model layers, except for the output layers, which undergo retraining on the new data. For this training dataset, we used 87 available PRG images and 900 non-PRG images from the catalog by \cite{Vavilova2021}. The pre-trained model with a new trainable output layer was trained over 35 epochs using the Adam optimizer with $10^{-4}$ learning rate. The model reached convergence, although there appears to be a discrepancy between training and validation loss. The model reached an unexpectedly high accuracy of 95.3\,\% on the test set, which could indicate overfitting.

\begin{figure}[h!] 
    \centering    \captionsetup{justification=centering,margin=1cm}
    \includegraphics[width=0.3\textwidth]{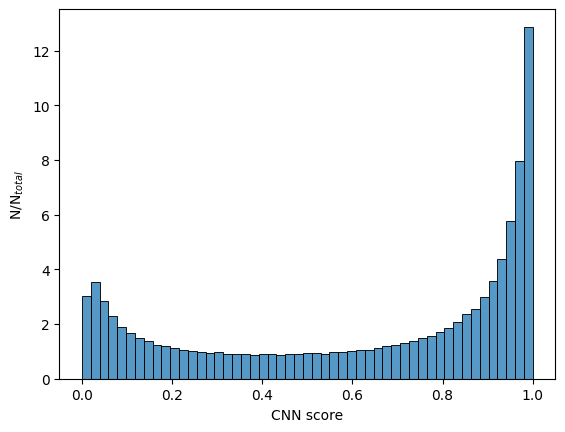}
    \caption{Distribution of CNN score for SDSS galaxies from catalog by \cite{Vavilova2021} to be PRG candidates.}
    \label{fig:CNN}
\end{figure}

Despite overfitting, we decided to apply the final model to the catalog of 315,000 SDSS galaxies by \cite{Vavilova2021}. \autoref{fig:CNN} shows the distribution of probabilities for galaxies to have a polar ring. A small number of examples for training led to the fact that the CNN ``remembered'' the pattern of galaxies with polar rings and could not work with a more complex sample with a large variety of morphological shapes of galaxies. In addition, the unevenness of the classes in the sample caused bias in the classification: the model often assigns either high or very high scores, ignoring the mean values. The CNN assigned a score of greater than 0.999 to a very large number of galaxies. So, 3,246 galaxies were enough, as expected due to the training data set imbalance in the number of PRGs and non-PRGs. 

\begin{figure*} [h!]
	\centering
	\includegraphics[width = 0.16\linewidth]{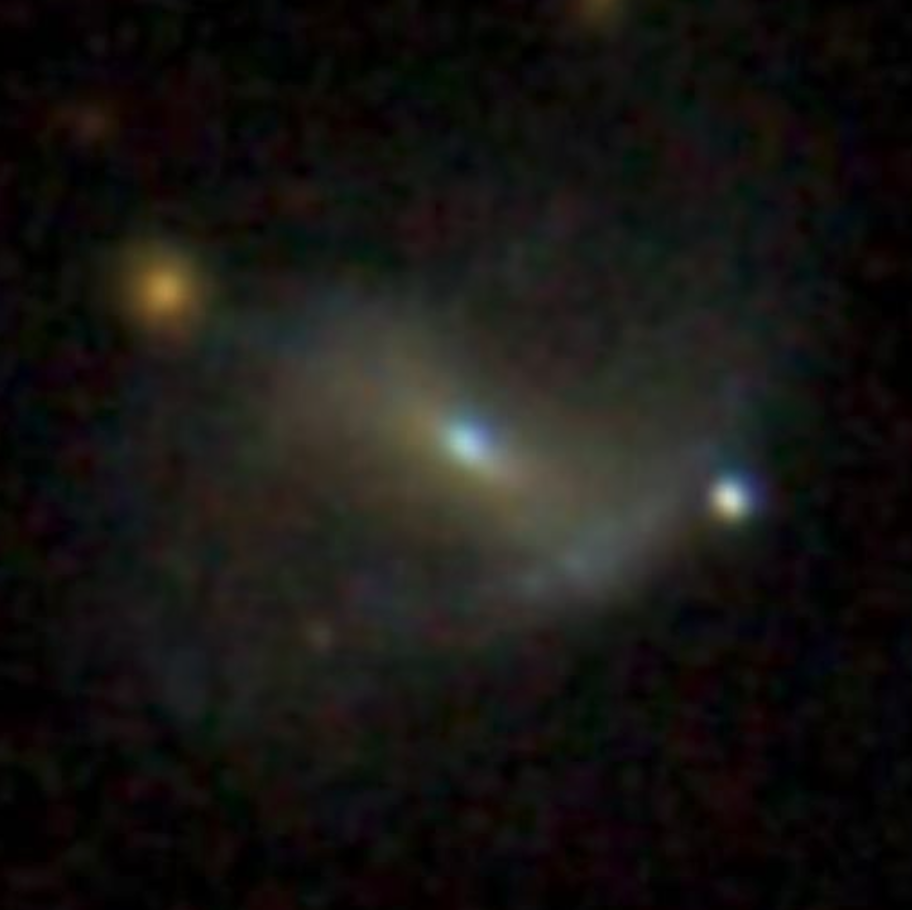}
    \includegraphics[width = 0.16\linewidth]{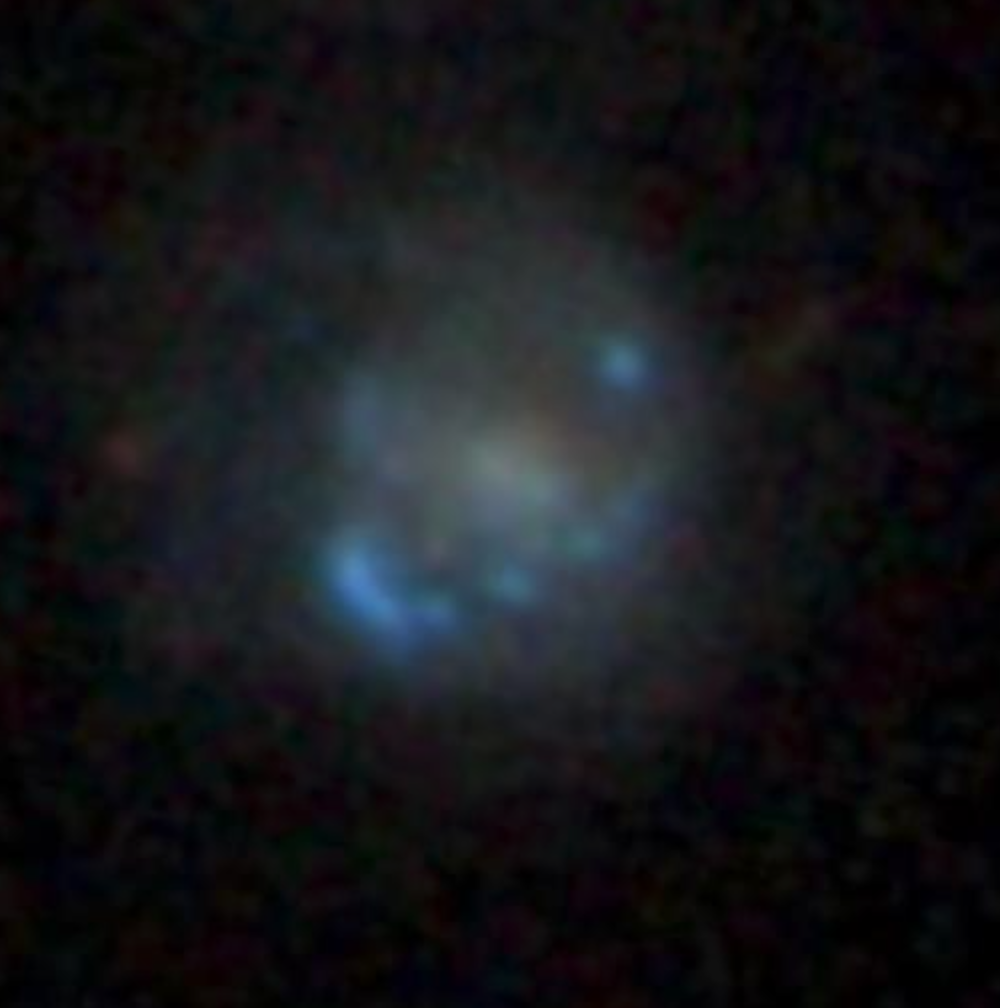}
    \includegraphics[width = 0.16\linewidth]{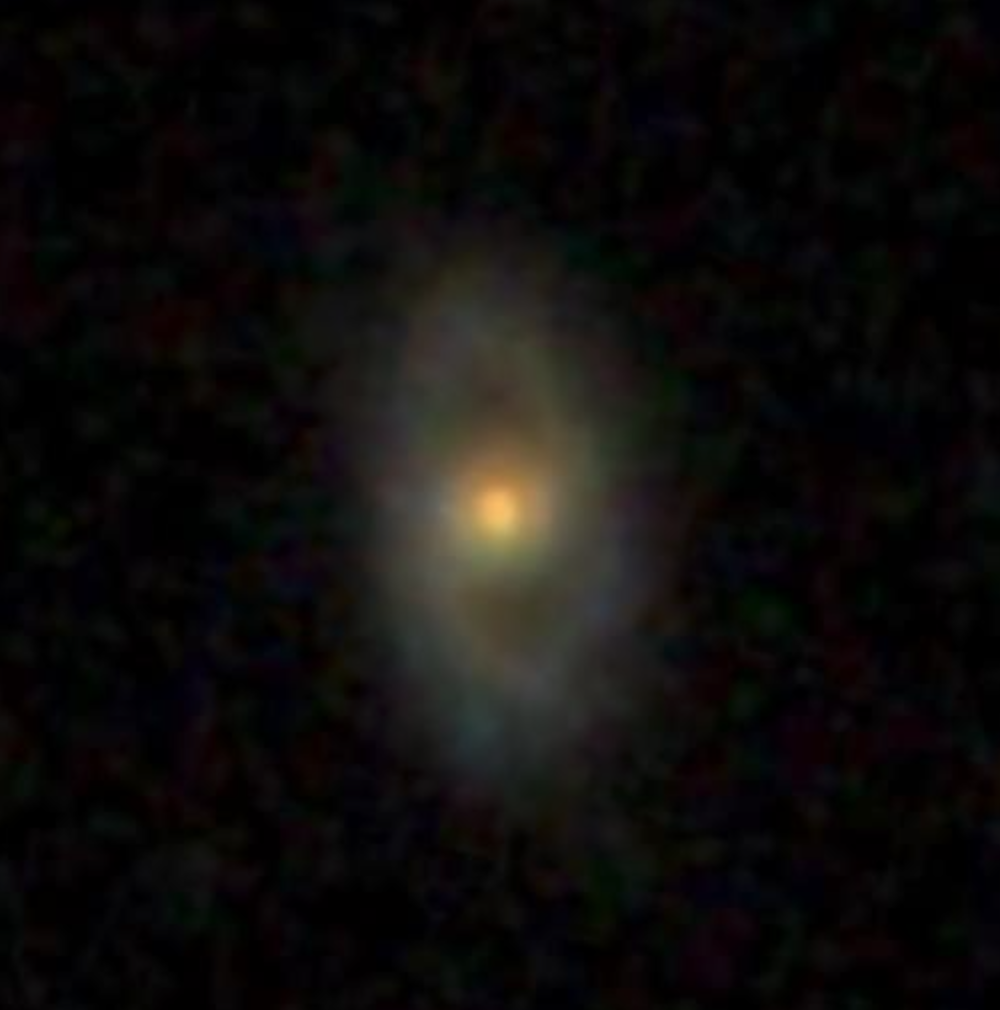}
    \includegraphics[width = 0.16\linewidth]{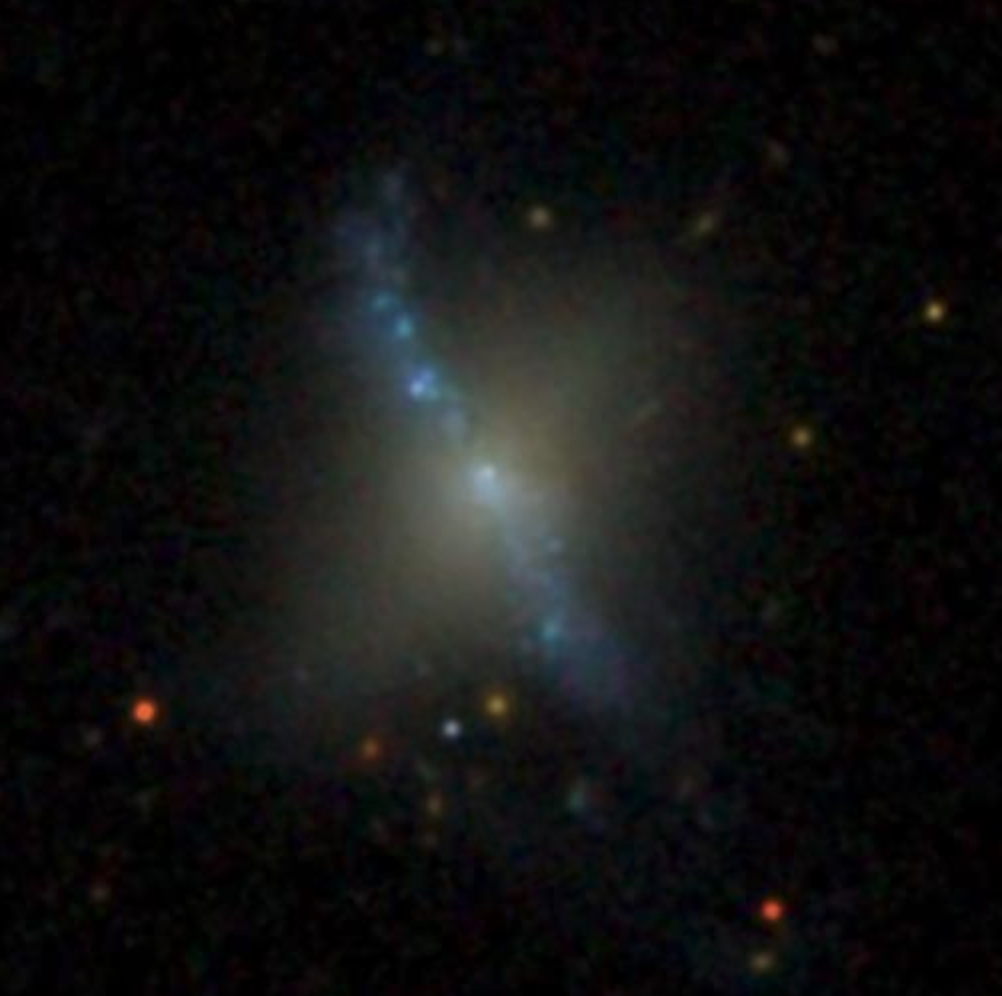}
    \includegraphics[width = 0.16\linewidth]{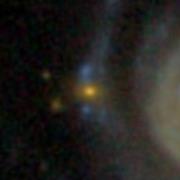}
	\caption{Examples of polar ring galaxy candidates identified by CNN: (a) SDSS, RA: 173.8703, DEC: 38.5550, z=0.022; (b) SDSS, RA: 151.3460, DEC: 19.2718, z=0.013; (c) SDSS, RA: 165.7470, DEC: 34.9276, z=0.032; (d) SDSS, RA: 222.8091, DEC: 35.5406, z=0.004  \citep{Reshetnikov2015}; (e) SDSS, RA: 195.6699, DEC: 50.4401, z=0.024.}
	\label{PRG_CNN_TL}
\end{figure*}

We have visually inspected these 3,246 galaxies that received a score greater than 0.999. Our experience with a machine learning photometry- and image- based approach for the classification of morphological features of galaxies from this SDSS catalog shows \citep{Vavilova2021, Vavilova2022, Khramtsov2022} that despite the probability of a galaxy having a specific feature being assigned by the CNN, visual inspection proves that the galaxy indeed has this feature, even if the CNN score is low. 

Our visual inspection pointed out that many objects with the highest scores appeared similar to objects that could be related to or reasonably confused with PRGs. The model selected many late spiral galaxies with loosely wound spiral arms, which may have been mistaken for rings. Additionally, CNN identified regular ring galaxies and merging galaxies that could be misinterpreted as having a ring-like structure. We describe below some interesting misclassified cases.

The images of late spirals are more precisely irregular-like, with the highest CNN score. We found two galaxies with coordinates RA: 173.8703, DEC: 38.5550, z=0.022, and RA: 151.3460, DEC: 19.2718, z=0.013 (`a and `b' in  \autoref{PRG_CNN_TL}, respectively). These images are similar to those of galaxies, which were classified as good candidates in the catalog by \cite{Reshetnikov2015} (see `d' in \autoref{PRG_CNN_TL}). Our CNN caught a few galaxies with coordinates RA: 222.8091, DEC: 35.5406, z=0.004 (see, `d' in \autoref{PRG_CNN_TL}), and RA: 161.5985, DEC: 6.6194, z=0.028, which are listed in the catalogs by \cite{Reshetnikov2015, Moiseev2011}. This means that the proposed CNN code was able to identify them.

Interesting example, when three galaxies were assigned as PRGs: 1) SDSS, RA: 195.6699, DEC: 50.4401, z=0.024; 2) SDSS, RA: 178.2528, DEC: 44.1393, z=0.070; 3) SDSS, RA: 3.9988, DEC: -0.3049, z=0.040. However, visual inspection and image zooming showed that they are elliptical galaxies located near a galaxy's spiral arm. Their overlap creates the appearance of rings around the elliptical (`e' in \autoref{PRG_CNN_TL}).

Our most significant finding in this study with transfer learning is the discovery of the SDSS J140644.42+471602.0 (see `a' in \autoref{SED}), which is a polar ring galaxy (SDSS, RA: 211.6850, DEC: 47.2672, z=0.060). The programming code of the CNN model with transfer learning, which allowed us to discover this PRG and reveal typical errors of CNN misclassification, is available on GitHub\footnote{\url{https://github.com/olhetm/PRG_classifier}}.

\section{Visually discovered PRGs from the Catalog of SDSS Ring galaxies}
\noindent
The Catalog of the SDSS Ring galaxies at $z$ < 0.1 is a part of a catalog developed with photometry- and image-based machine learning techniques \citep{Vavilova2023}. We have done a visual inspection of 13,882 objects, which allowed us to identify and remove misclassified machine learning objects and enhance the accuracy of the morphological classification of the ring feature. The Catalog of the SDSS Ring galaxies, containing $\sim$2,200 objects, is available on the UkrVO website\footnote{\url{https://www.mao.kiev.ua/index.php/ua/vpaai-labvmsv-catalogues}}.

\begin{figure*} [h!]
	\centering
    \includegraphics[width = 0.20\linewidth]{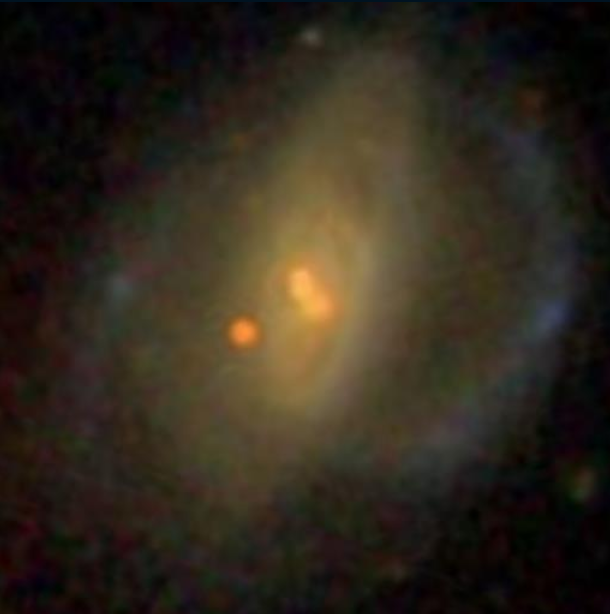}
	\includegraphics[width = 0.20\linewidth]{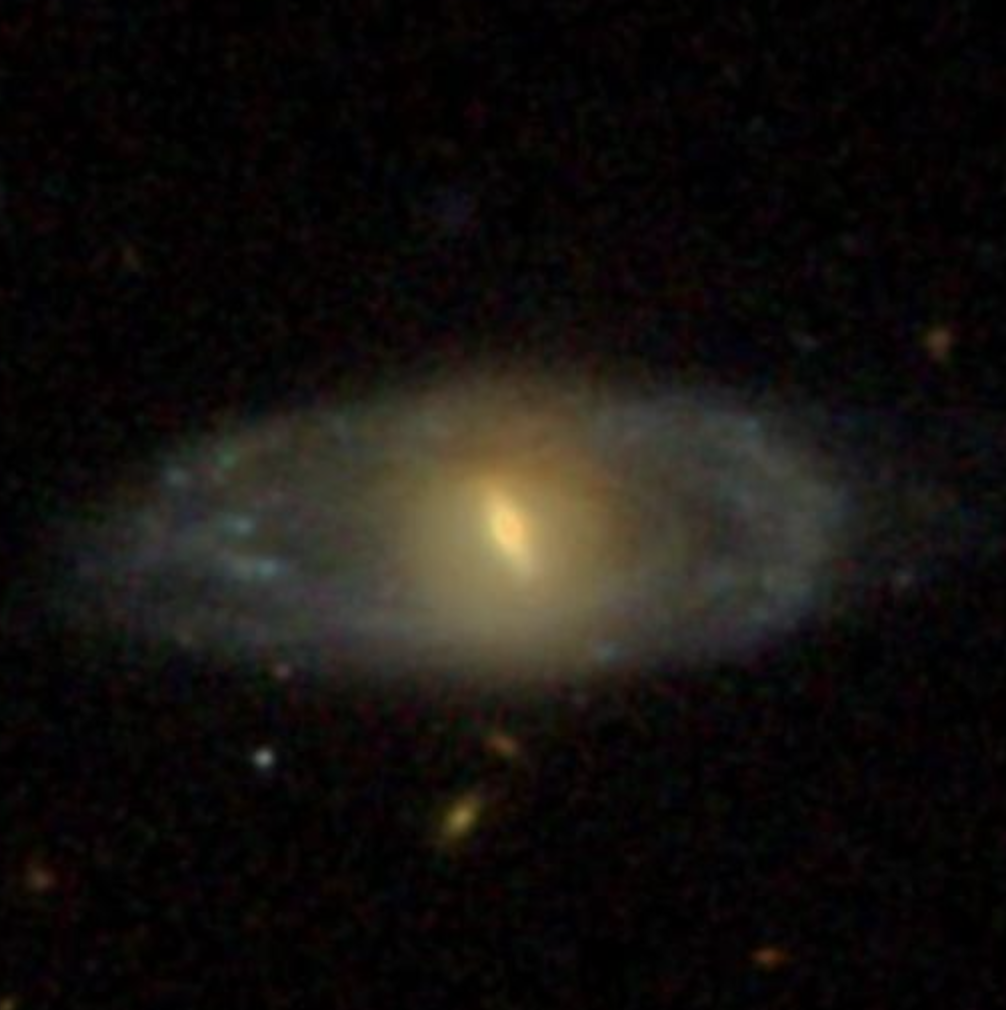}
    \includegraphics[width = 0.20\linewidth]{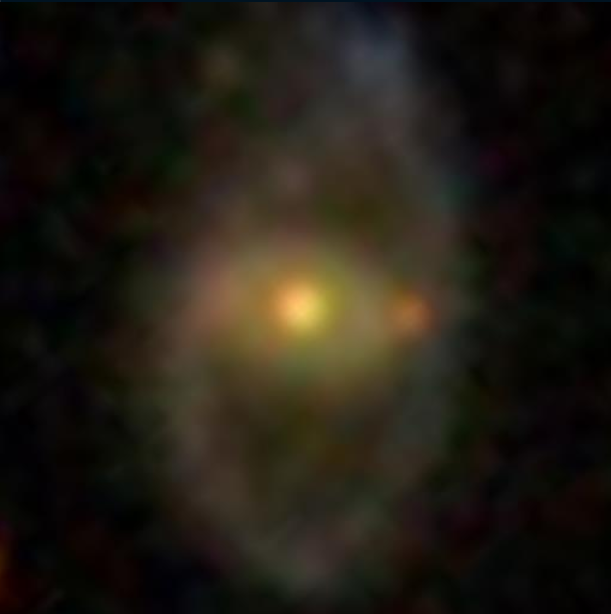}
    \includegraphics[width = 0.20\linewidth]{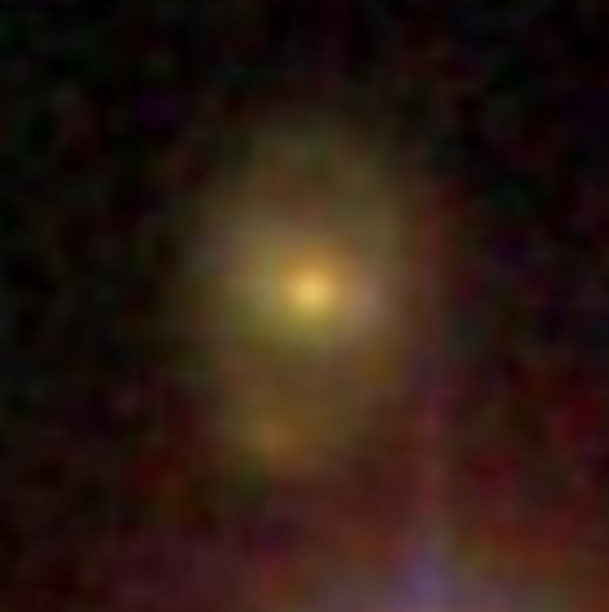}
	\caption{Discovered PRGs from Catalog of SDSS Ring galaxies after visual inspection: (a) RA: 149.7138, DEC: 32.0730, z=0.027; (b) RA: 160.5460, DEC: 23.7467, z=0.012; (c) RA: 245.5526, DEC: 27.3422, z=0.096; (d) RA: 161.5004, DEC: 9.1075, z=0.087.}
	\label{PRG_R_VI_cat_PRG}
\end{figure*}

\begin{figure*} [h!]
	\centering
    \includegraphics[width = 0.20\linewidth]{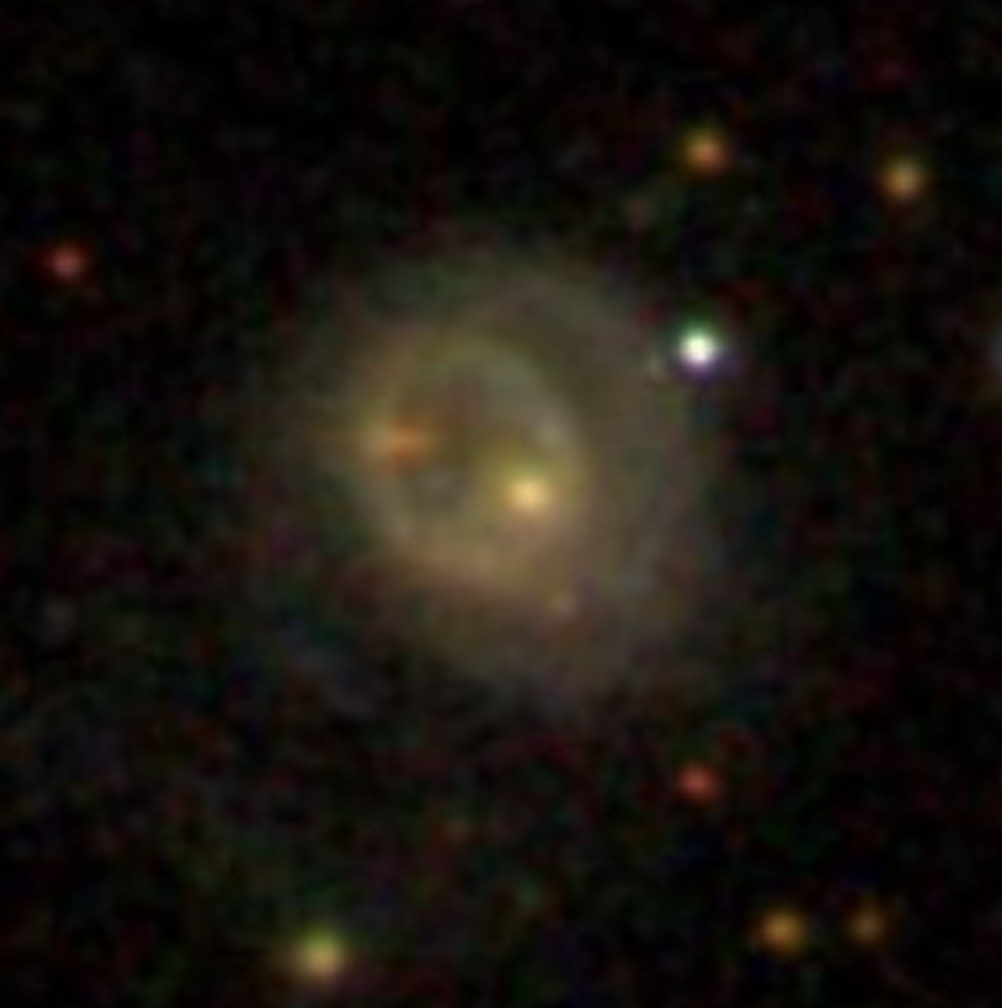}
	\includegraphics[width = 0.20\linewidth]{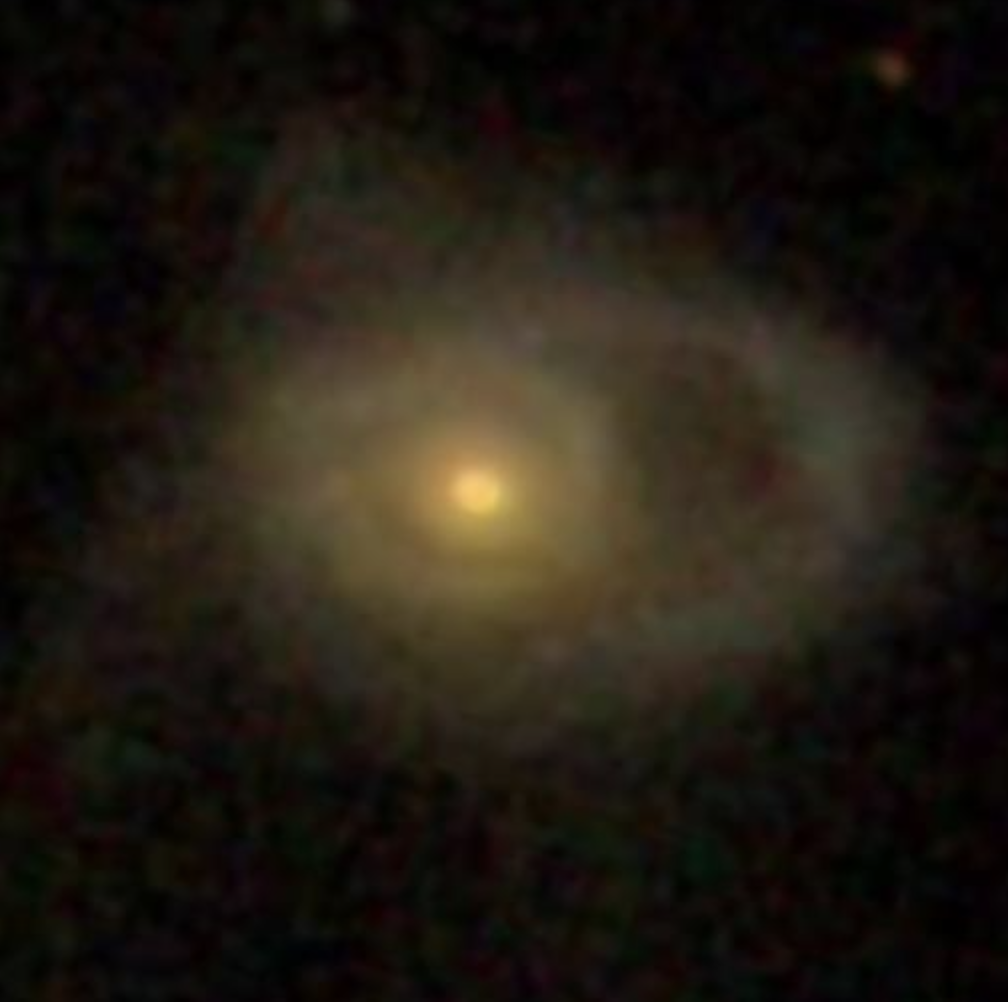}
    \includegraphics[width = 0.20\linewidth]{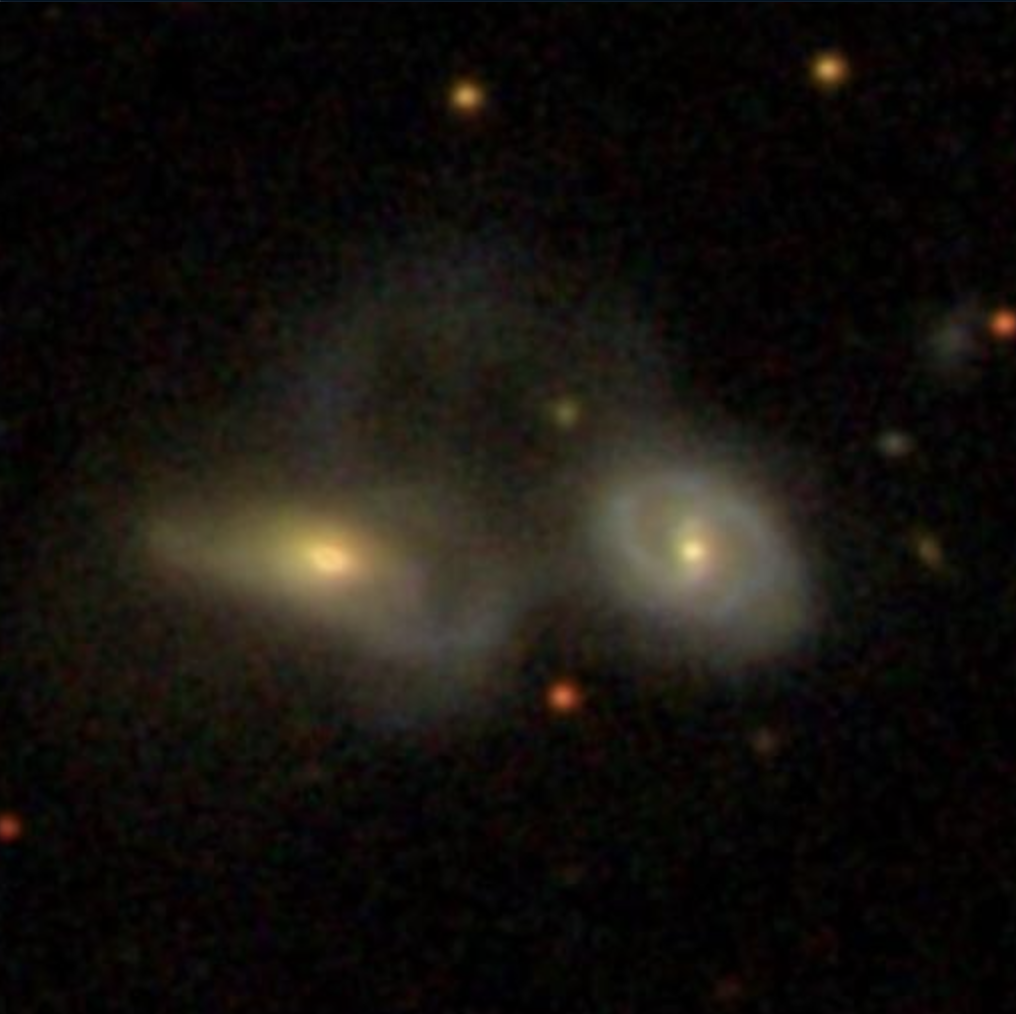}
    \includegraphics[width = 0.20\linewidth]{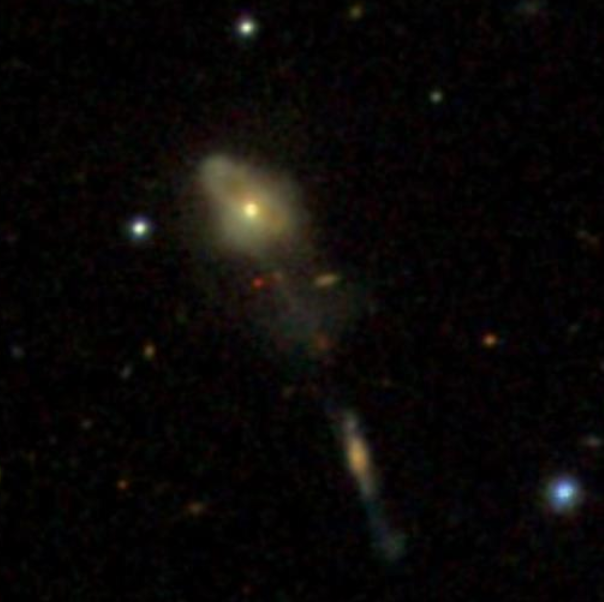}
	\caption{Examples of double ring galaxies from the Catalog of Ring galaxies by visual inspection (a) RA: 227.7441, DEC: 5.3180, z=0.081; (b) RA: 138.8240, DEC: 18.0789, z=0.056. Examples of galaxies with ring formation through merging from the Catalog of Ring galaxies by visual inspection: (c) RA: 198.8077, DEC: 44.4073, z=0.035; (d) RA: 198.6566, DEC: 26.1239, z=0.073.}
	\label{PRG_R_VI_cat}
\end{figure*}

During the visual inspection of $\sim$2,200 ring galaxies, we discovered four PRG candidates: (a) SDSS, RA: 149.7138, DEC: 32.0730, z=0.027; (b) SDSS, RA: 160.5460, DEC: 23.7467, z=0.012; (c) SDSS, RA: 245.5526, DEC: 27.3422, z=0.096; (d) SDSS, RA: 161.5004, DEC: 9.1075, z=0.087. Their images are presented in \autoref{PRG_R_VI_cat_PRG}. Interesting examples of galaxies are like double rings (`a' and `b' in \autoref{PRG_R_VI_cat}) as well as systems, where the formation of a polar ring may be underway (`c' and `d' in \autoref{PRG_R_VI_cat}).

\section{Multiwavelength analysis of the discovered PRG SDSS J140644.42+471602.0}
\noindent
\subsection{Data}

\begin{figure*} [h!]
	\centering
	\includegraphics[width = 0.40\linewidth]{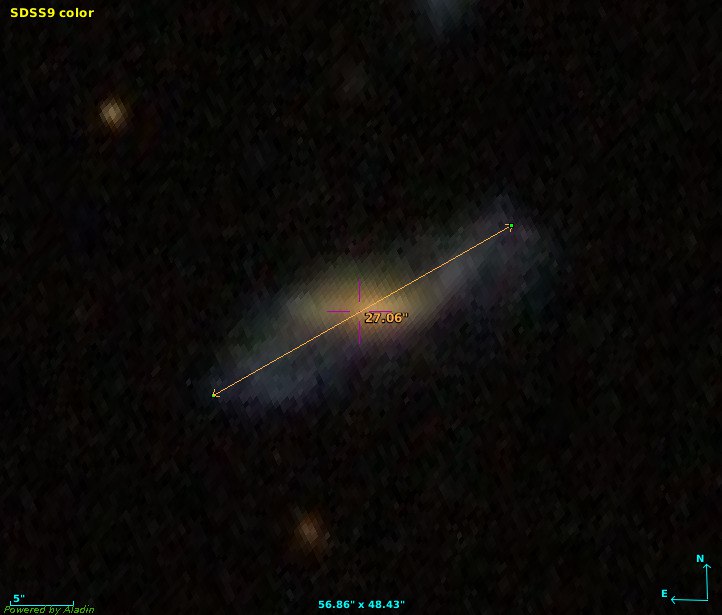}
    \includegraphics[width = 0.50\linewidth]{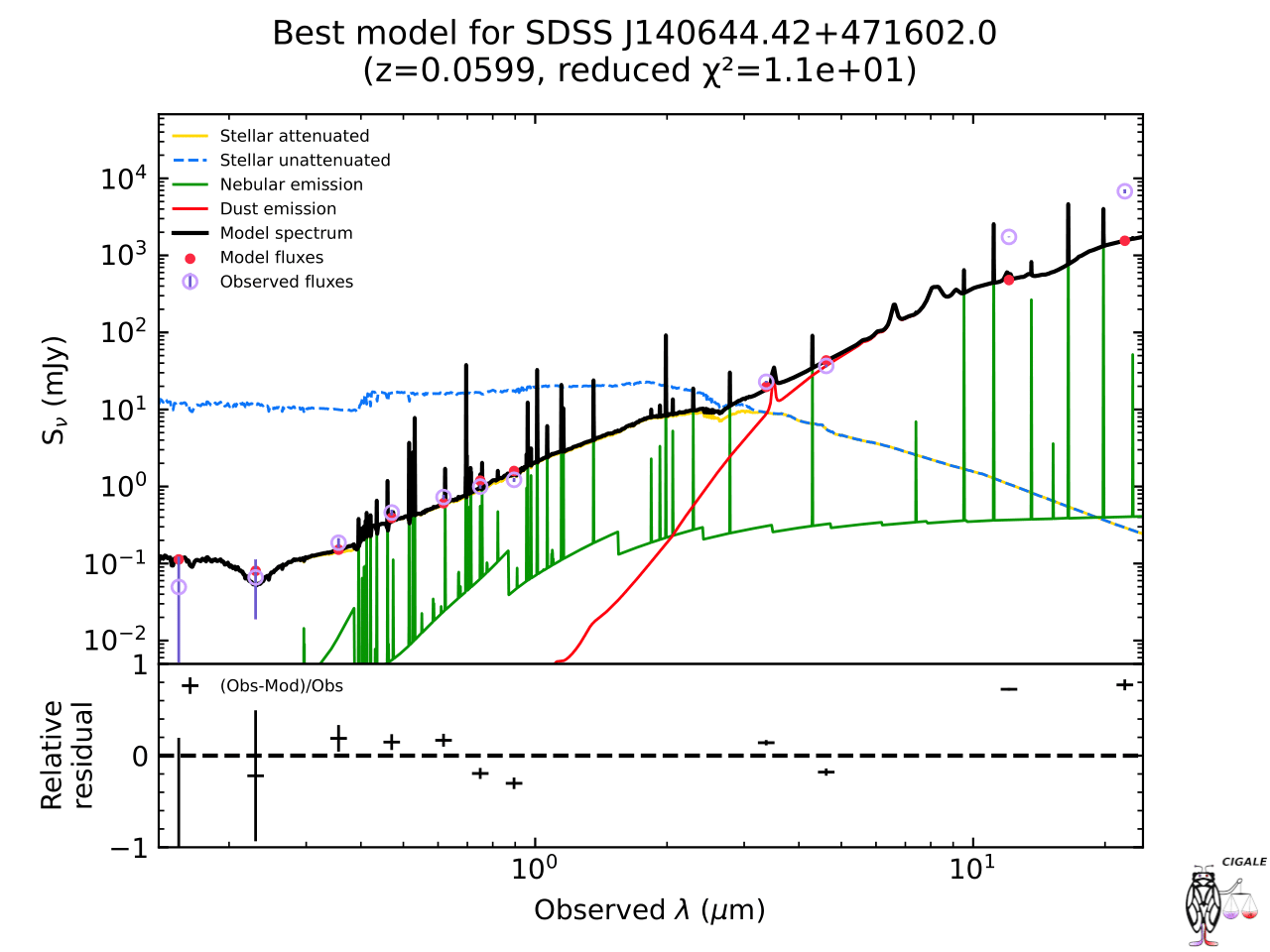}
	\caption{(a) PRG SDSS J140644.42+471602.0 discovered with transfer learning. The yellow line is the estimated line of sight size in the optical band. The image is obtained by the Aladin tool. (b)  The best approximation of the SED from UV to IR bands. Red dots are model fluxes; purple dots are observed fluxes. The green curve describes the nebular emission, the yellow curve - the stellar component without attenuation, the blue dashed curve describes the stellar component with attenuation, the red curve - dust radiation. The black curve is the resulting model SED.}
	\label{SED}
\end{figure*}

The initial data for the galaxy SDSS J140644.42+471602.0 was obtained from the SDSS survey. However, data across the entire electromagnetic spectrum was required for a comprehensive multiwavelength analysis. We found that this galaxy was also observed by the GALEX and WISE space observatories. The standard circular apertures used for photometric measurements in these surveys differ in size: GALEX (UV) employs diameters of 12 arcsec for FUV and 34.5 arcsec for NUV bands; SDSS (optical) applies a 6-arcsecond aperture for all bands; WISE (IR) uses 16.5 arcsec for the W1, W2, and W3 bands and 33 arcsec for W4. The total size of the galaxy, including its ring, is $\approx$27 arcsec (see`a' in \autoref{SED}).

The preprocessed, cleaned, and calibrated UV (GALEX) and IR (WISE) images of the galaxy were downloaded from the IRSA archive\footnote{https://irsa.ipac.caltech.edu/Missions/wise.html} and the optical (SDSS) images from the SDSS. To further analyze these data, we used the Aperture Photometry Tool (APT) v. 3.0.8\footnote{\url{https://www.aperturephotometry.org}} \cite{Laher2012} to derive the magnitudes from these images accurately. Since the observed size of the galaxy varies across the wavelength bands, we used the radial profile and curve of growth to determine the optimal aperture size for source and background, as recommended in the APT guide. The final photometric values ensuring accurate background subtraction were obtained with the Sky average subtraction model, and more details of all available models can be found in the work by \cite{Laher2012}. 

For further multiwavelength analysis (subsection 5.2), we consulted with the SDSS team to consider that SDSS magnitudes are asinh magnitudes and are related to flux through the inverse hyperbolic sine function. We encountered completely unphysical magnitude values when applying the Pogson equation with a zero point of 22.5. The resulting fluxes from UV to IR for the entire galaxy with a ring are given in \autoref{tab:FLUX}: two UV bands (FUV and NUV); five SDSS bands ($u,g,r,i,z$), and four IR bands (W1, W2, W3, and W4).  Galactic reddening and extinction were considered a k-correction based on the IRSA dust map\footnote{https://irsa.ipac.caltech.edu/applications/DUST/}. 

\subsection{Spectral energy distribution for SDSS J140644.42+471602.0}

\begin{table*}[h!]
    \centering
     \caption{The UV to IR k-corrected observed fluxes in mJy for SDSS J140644.42+471602.0}
\begin{tabular}{|c|c|c|c|c|c|c|c|c|c|c|}
\hline
  FUV & FUV\_err & NUV & NUV\_err & u & u\_err & g & g\_err & r & r\_err & i \\
\hline
  0.05 & 0.07 & 0.07 & 0.05 & 0.19 & 0.03 & 0.46 & 0.04 & 0.73 & 0.05 & 1.0 \\
\hline
  i\_err & z & z\_err & W1 & W1\_err & W2 & W2\_err & W3 & W3\_err & W4 & W4\_err \\
\hline
  0.06 & 1.22 & 0.08 & 22.88 & 0.66 & 36.74 & 1.44 & 1738.71 & 22.10 & 6801.83 & 405.33 \\
\hline
\end{tabular}
    \label{tab:FLUX}
\end{table*}

Galaxies with polar rings demonstrate a complex star formation history influenced by the formation of the central galaxy's main stellar population and the subsequent formation of the polar ring. As a result, we have varying ages: an older population dominates the central galaxy, while a younger population is typically found in the polar ring.

To obtain more detailed properties of SDSS J140644.42+471602, we exploited CIGALE software \citep{Boquien2019} for multiwavelength analysis of spectral energy distribution (SED) from UV to IR spectral ranges with the following modules:
\begin{description}
   \item[\textit{Star formation history}] was described with \textit{sfh2exp} module, which has two exponential components. The first allows us to describe the early stages of galaxy evolution and the initial burst of star formation. The second exponential component is responsible for star formation at the later evolutionary stages. 
    \item[\textit{Stellar population}] was described with \textit{bc03} module \cite{Bruzual2003}. 
    \item[\textit{Nebular emission}] with \textit{nebular} module simulates the ionisation spectrum that arises from young hot stars using results from stellar population model by \cite{Inoue2011}. 
    \item[\textit{Dust attenuation}] was taken into account with \textit{dustatt$\_$modified$\_$starburst} module, based on the modified attenuation law \citep{Calzetti2000, Leitherer2002}.
    It uses modified absorption curves for different galaxy components, including young and old stellar populations. 
    \item[\textit{Dust emission}] with \textit{dl2014} module describes the heating and radiation of interstellar dust in galaxies \citep{Draine2014}.  
\end{description}
      
The Bayesian method selected the best model for the observed SED (\autoref{SED}) from more than 1 billion possible combinations. The history of star formation is described by two exponential flares, as mentioned above. The E-folding time of the main stellar population is 500 Myr, which led to the formation of the stellar population in the central galaxy with a mass of M$_{star}^{old}$ = 5.73$\times 10^{10}$ $M_{\odot}$ and an age of 8.5 Gyr. The second episode of star formation began 500 Myr ago with an E-folding time of 4 Gyr, leading to the formation of a younger stellar population, M$_{star}^{young}$ = 2.6$\times 10^{10}$ $M_{\odot}$. 

The estimated E-folding time of the late stellar population suggests that the ring formed due to accretion from a satellite galaxy. This most likely led to its destruction, since the galaxy has no visible neighbors on the scale of 0.5 angular seconds. We also assume here that the formation of a polar ring resulted from a merger. Numerical simulations by \cite{Matteo2008} show that in the case of a merger, the E-folding time of the late stellar population should be less than 1 Gyr, and variants with larger values arise in the case of accretion. Now, the formation of new stars continues, consistent with the classification of this galaxy in the SDSS based on spectral data. The value of SFR in our model is 71 $M_{\odot}$ per year. This is only a rough estimate. Due to the lack of observational data in the FUV band, we cannot correctly estimate star formation since the radiation from young stars surrounded by a dust cocoon is not considered. The estimated total mass of the stellar component is M$_{star}$ = 8.34$\times 10^{10}$ $M_{\odot}$ and the dust mass based on this model is M$_{dust}$ = 5$\times 10^{6}$ $M_{\odot}$.

\section{Summary and conclusions}
\noindent
\subsection{Catalog of Inspected Polar Ring Galaxies}

Our Catalog of Inspected PRGs consists of 179 objects:

- 167 objects that cover the categories of strong, good, and weak candidates based on the data by \citep{Whitmore1990, Moiseev2011, Reshetnikov2019}, which have high-quality SDSS images;

- three objects found by \cite{Skryabina2024}, one PRG discovered by \cite{Akhil2024b} and one PRG confirmed by \cite{Freitas2024};

- three PRGs discovered by us with a deep learning approach (sections 3 and 4) and four PRGs found visually from the Catalog of SDSS Ring Galaxies (section 5). 

The preliminary PRG sample (Section 2) also includes 108 objects of the southern sky, and their validation is in progress.  

\begin{figure}[h!] 
    \centering    \captionsetup{justification=centering,margin=1cm}
    \includegraphics[width=0.5\textwidth]{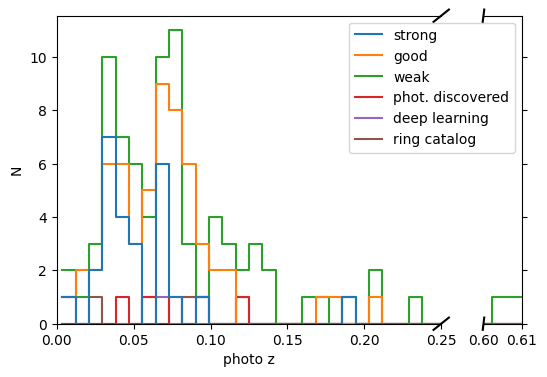}
    \caption{Distribution of inspected PRGs by redshift }
    \label{fig:histogr}
\end{figure}

\begin{figure}[h!] 
    \centering    \captionsetup{justification=centering,margin=1cm}
    \includegraphics[width=0.5\textwidth]{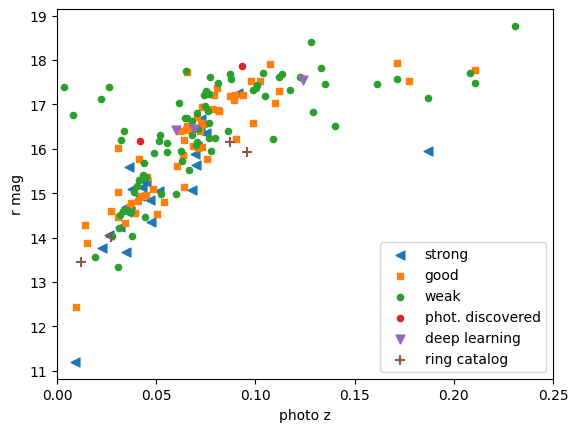}
    \caption{Dependence of magnitude vs. redshift in $r$-filter for inspected PRGs}
    \label{fig:rmag}
\end{figure}

\autoref{fig:histogr} and \autoref{fig:rmag} demonstrate the distribution of these PRGs by redshift located mostly in the Local Volume as well as the dependence of magnitude vs. redshift in $r$-filter, respectively. This Catalog will be supplemented in VizieR.

We also cross-matched inspected PRGs to have AGNs like \cite{Smirnov2020}, who considered the frequency of AGNs among galaxies with large-scale optical polar structures based on SDSS spectra. The BPT diagrams led to the conclusion that among the studied PRGs, there is a possible excess of AGNs, which needs further confirmation.

\subsection{Multiwavelength analysis}

Data availability across different electromagnetic spectrum ranges enables multiwavelength analysis using the CIGALE program. However, there are no observations in the far-infrared and radio bands for most newly identified PRGs. This limitation affects the accurate estimation of the star formation rate in these galaxies, as the absence of such data prevents a proper accounting of contributions from star-forming regions that may be embedded in dust. Nevertheless, we found that 17 studied PRGs exhibit a significant scatter in their neutral hydrogen content relative to the stellar mass, with HI mass ranging from 0.063 $\%$ to 63$\%$ of the stellar mass across different galaxies. Some of these galaxies appear very bright on UV images from the GALEX space observatory, indicating a large amount of gas and potentially active star formation.

Here, we test a new approach for multiwavelength analysis of PRGs. Since line-of-sight sizes are bigger than standard aperture sizes for different space and ground observatories, we used the Aperture Photometry Tool (APT) v. 3.0.8 \cite{Laher2012} to obtain accurate photometry in each band. We have analyzed a new PRG, SDSS J140644.42+471602.0, with CIGALE software. One of the main challenges was the inability to separate the disk and ring components, which complicates the search for the optimal star formation model. However, the model techniques helped identify the most plausible evolutionary scenarios for forming the polar ring based on the E-folding time of the late-burst stellar population. The results suggest that the galaxy’s stellar population formed in two main episodes: an older population with an e-folding time of 500 Myr and a younger population formed 500 Myr ago with an e-folding time of 4 Gyr. The formation of the polar ring is most likely due to accretion from a satellite galaxy rather than a merger.  The current SFR for this PRG is 71 $M_{\odot}$ per year, although the lack of FUV data limits this estimate. The total stellar mass is estimated to be 8.34$\times 10^{10}$ $M_{\odot}$, with a dust mass of 5$\times 10^{6}$ $M_{\odot}$. The predominance of an old stellar population (two-thirds of the total mass) suggests that this PRG is undergoing interaction processes (see, for example, \cite{Freitas2024} for ESO 287-IG50). 

\subsection{Deep learning approach for PRG search}

For the first time, we applied machine learning image-based methods to the PRG search even though only about 463 PRG candidates are currently known. This risk turned out to be justified: we discovered three PRGs. Because we can not compare our results with others, we will discuss and summarize a methodology of our approach. 

We examined the SDSS sample of high-quality galaxy images; nevertheless, that training sample was extremely limited by 87 PRGs (we did not include even weak PRG candidates because we wanted the neural network to operate with well-defined images). Moreover, we expected that training on such a small sample of images would likely lead to overfitting. Therefore, developing a functional neural network approach for this problem requires employing various techniques to address the significant class imbalance. 

We tried to solve this problem in several ways. We need to increase the training sample. Firstly, we used augmentation, which gave a good result in our previous works \citep{Khramtsov2022}. we applied PyTorch (\cite{Paszke2019}), which has built-in augmentation functions. This technique artificially increases the size of the data set by generating different versions of the original images using operations such as rotation, flipping, scaling, and color adjustment. These additions help the model learn more generalized features, thereby improving its ability to recognize PRGs in various contexts. In addition, we used image segmentation, which helped isolate the regions of interest and concentrate on the PRGs. This approach reduces noise and improves the accuracy of the feature extraction process. Furthermore, we implemented an ensemble of models to boost classification performance. Ensemble learning involves combining the predictions of multiple models to produce a final prediction. This technique mitigates the risk of relying on a single model, which might be prone to overfitting or underfitting. For example, we achieved a more balanced and reliable classification outcome by averaging the predictions from various models. In the result of this ensemble approach at the first stage, 494 galaxies from $\approx$ 315,000 were classified as PRG with scores higher than 0.9. However, upon visual inspection, none of these galaxies exhibited the characteristics of PRGs.

Another independent approach was that a shallow neural network with one hidden layer was trained, where we also used augmentation but slightly differently; PRG images were increased by 25 times through data augmentation. When applied to a representative SDSS dataset, the model made a large number of false-positive predictions. We selected 25 potential candidates for visual inspection with an accuracy of 0.94 to be PRG or related objects. As a result, we have found a pair of galaxies (second row in \autoref{PRG_A}) with a ring pattern. Since they are fairly closed, this pair should be investigated in more detail. If a more detailed analysis confirms that there is a ring in this pair (second row in \autoref{PRG_A}), this option can be considered as additional information for the formation of PRGs due to the merger process \citep{Schweizer1983, Bekki1997, Bekki1998, Akhil2024b} or by the gas action of the donor galaxy on the dominant galaxy \cite{Bournaud2003, Ordenes2016}. 
\cite{Freitas2024} explored the peculiar galaxy ESO 287-IG50 (PRG candidate from the Atlas by \cite{Whitmore1990}. Their photometry and spectroscopic analysis led to the conclusion that this PRG is in an ongoing formation process

As a final method to solve the problem of insufficient training sample, we attempted transfer learning -- training using synthetic images generated via GALFIT \citep{Peng2002}. \cite{Krishnakumar2022} utilized transfer learning to train a CNN to identify ring galaxies. They used a modified GalSim code \cite{Ghosh2020} to generate a sample of simulated ring galaxies and then run GALFIT, allowing for an automatic generation of the required number of images. We also followed a similar path and created a chosen number of GALFIT input files with the simulation parameters. Then, GALFIT is run on these files, producing the simulated images. Lastly, the noise is added to the images, and conversion from FITS to JPG is done. 

Our model, again, overfit to the data during the training process, as in the previous run, but we decided to test the final model on a \cite{Vavilova2021} sample with more than 300k galaxies from SDSS. We have visually inspected galaxies with an accuracy of 0.999 and found interesting results. The first promising sign is that the model selected a lot of late spiral galaxies with tight spiral arms, which the model could probably confuse with rings. Secondly, the model chose a lot of regular ring galaxies and merging galaxies, which could be confused with having a ring-like structure. 

So, our deep learning approach has resulted in the discovery of three PRGs (RA: 211.6850, DEC: 47.2672, z=0.060; RA: 204.2103, DEC: 49.4625, z=0.097; RA: 149.3220, DEC: 36.8315, z=0.053) as well as four PRGs (RA: 149.7138, DEC: 32.0730, z=0.027; RA: 160.5460, DEC: 23.7467, z=0.012; RA: 245.5526, DEC: 27.3422, z=0.096; RA: 161.5004, DEC: 9.1075, z=0.087) were discovered upon visual inspection of Catalog of the SDSS Ring Galaxies. 

Our outcome suggests that deep learning is a promising way for PRGs search and requires some further refinement and potential enhancements to the developed methodology. Incorporating of some additional information (photometry or spectroscopic data) will enable a hybrid analysis (see, for example, \cite{Gutirrez2011}). Another promising way is to generate new training samples using contemporary Generative Adversarial Networks (see, for example, \cite{Goodfellow2014, Vavilova2018}). The GANs can create realistic synthetic data for wider existing dataset, thereby dealing with the issue of limited and imbalanced data and helping with the generalization and robustness of the model. In both cases the Cataog of inspected PRG images being supplemented with new discovered objects will become quite useful as for CNN approach and theoretical stidies. These strategies represent valuable opportunities for future development, aiming to push the boundaries of our current model and achieve more reliable identification of PRGs in big sky surveys.

\begin{acknowledgements}

The research by Dobrycheva D.V., Hetmantsev O.O., Vavilova I.B. and Kompaniiets O.V. is part of the project (ID 848), which is conducted in the frame of the EURIZON program funded by the European Union under grant agreement No.871072. We thank Prof. J. Knapen and Prof. J. Beckman (IAC, Spain) and Prof. M. Kowalski (DESY, Germany) for fruitful seminars and discussions in March 2025, when the results of this research were finalized.

\end{acknowledgements}

\bibliographystyle{aa} 
\bibliography{PRG_AA} 

\begin{thebibliography}{66}
\expandafter\ifx\csname natexlab\endcsname\relax\def\natexlab#1{#1}\fi

\bibitem[{{Akhil} {et~al.}(2025){Akhil}, {Kartha}, {Krishnan}, {Mathew},
  {Thomas}, {Ray}, \& {Devaraj}}]{Akhil2025}
{Akhil}, K.~R., {Kartha}, S.~S., {Krishnan}, U., {et~al.} 2025, arXiv e-prints,
  arXiv:2503.03709

\bibitem[{{Akhil} {et~al.}(2024{\natexlab{a}}){Akhil}, {Kartha}, \&
  {Mathew}}]{Akhil2024b}
{Akhil}, K.~R., {Kartha}, S.~S., \& {Mathew}, B. 2024{\natexlab{a}}, \mnras,
  530, 2907

\bibitem[{{Akhil} {et~al.}(2024{\natexlab{b}}){Akhil}, {Kartha}, {Mathew},
  {Ujjwal}, {Ezhikode}, \& {Robin}}]{Akhil2024}
{Akhil}, K.~R., {Kartha}, S.~S., {Mathew}, B., {et~al.} 2024{\natexlab{b}},
  \aap, 681, A35

\bibitem[{{Arp} \& {Madore}(1987)}]{Arp1987}
{Arp}, H.~C. \& {Madore}, B. 1987, {A catalogue of southern peculiar galaxies
  and associations}

\bibitem[{{Bahr} \& {Mosenkov}(2024)}]{Bahr2024}
{Bahr}, S. \& {Mosenkov}, A. 2024, in American Astronomical Society Meeting
  Abstracts, Vol.~56, American Astronomical Society Meeting Abstracts, 408.04

\bibitem[{{Bekki}(1997)}]{Bekki1997}
{Bekki}, K. 1997, \apjl, 490, L37

\bibitem[{{Bekki}(1998)}]{Bekki1998}
{Bekki}, K. 1998, \apj, 499, 635

\bibitem[{{Beygu} {et~al.}(2011){Beygu}, {Jarrett}, {Jarrett}, {van de
  Weygaert}, {Kreckel}, {van der Hulst}, \& {van Gorkom}}]{Beygu2011}
{Beygu}, B., {Jarrett}, T., {Jarrett}, T., {et~al.} 2011, Spitzer Proposal ID
  80069, 80069

\bibitem[{{Blanton} {et~al.}(2017){Blanton}, {Bershady}, {Abolfathi},
  {Albareti}, {Allende Prieto}, {Almeida}, {Alonso-Garc{\'\i}a}, {Anders},
  {Anderson}, {Andrews}, {Aquino-Ort{\'\i}z}, {Arag{\'o}n-Salamanca},
  {Argudo-Fern{\'a}ndez}, {Armengaud}, {Aubourg}, {Avila-Reese}, {Badenes},
  {Bailey}, {Barger}, {Barrera-Ballesteros}, {Bartosz}, {Bates}, {Baumgarten},
  {Bautista}, {Beaton}, {Beers}, {Belfiore}, {Bender}, {Berlind}, {Bernardi},
  {Beutler}, {Bird}, {Bizyaev}, {Blanc}, {Blomqvist}, {Bolton}, {Boquien},
  {Borissova}, {van den Bosch}, {Bovy}, {Brandt}, {Brinkmann}, {Brownstein},
  {Bundy}, {Burgasser}, {Burtin}, {Busca}, {Cappellari}, {Delgado Carigi},
  {Carlberg}, {Carnero Rosell}, {Carrera}, {Chanover}, {Cherinka}, {Cheung},
  {G{\'o}mez Maqueo Chew}, {Chiappini}, {Choi}, {Chojnowski}, {Chuang},
  {Chung}, {Cirolini}, {Clerc}, {Cohen}, {Comparat}, {da Costa}, {Cousinou},
  {Covey}, {Crane}, {Croft}, {Cruz-Gonzalez}, {Garrido Cuadra}, {Cunha},
  {Damke}, {Darling}, {Davies}, {Dawson}, {de la Macorra}, {Dell'Agli}, {De
  Lee}, {Delubac}, {Di Mille}, {Diamond-Stanic}, {Cano-D{\'\i}az}, {Donor},
  {Downes}, {Drory}, {du Mas des Bourboux}, {Duckworth}, {Dwelly}, {Dyer},
  {Ebelke}, {Eigenbrot}, {Eisenstein}, {Emsellem}, {Eracleous}, {Escoffier},
  {Evans}, {Fan}, {Fern{\'a}ndez-Alvar}, {Fernandez-Trincado}, {Feuillet},
  {Finoguenov}, {Fleming}, {Font-Ribera}, {Fredrickson}, {Freischlad},
  {Frinchaboy}, {Fuentes}, {Galbany}, {Garcia-Dias},
  {Garc{\'\i}a-Hern{\'a}ndez}, {Gaulme}, {Geisler}, {Gelfand},
  {Gil-Mar{\'\i}n}, {Gillespie}, {Goddard}, {Gonzalez-Perez}, {Grabowski},
  {Green}, {Grier}, {Gunn}, {Guo}, {Guy}, {Hagen}, {Hahn}, {Hall}, {Harding},
  {Hasselquist}, {Hawley}, {Hearty}, {Gonzalez Hern{\'a}ndez}, {Ho}, {Hogg},
  {Holley-Bockelmann}, {Holtzman}, {Holzer}, {Huehnerhoff}, {Hutchinson},
  {Hwang}, {Ibarra-Medel}, {da Silva Ilha}, {Ivans}, {Ivory}, {Jackson},
  {Jensen}, {Johnson}, {Jones}, {J{\"o}nsson}, {Jullo}, {Kamble}, {Kinemuchi},
  {Kirkby}, {Kitaura}, {Klaene}, {Knapp}, {Kneib}, {Kollmeier}, {Lacerna},
  {Lane}, {Lang}, {Law}, {Lazarz}, {Lee}, {Le Goff}, {Liang}, {Li}, {Li},
  {Lian}, {Lima}, {Lin}, {Lin}, {Bertran de Lis}, {Liu}, {de Icaza Lizaola},
  {Long}, {Lucatello}, {Lundgren}, {MacDonald}, {Deconto Machado}, {MacLeod},
  {Mahadevan}, {Geimba Maia}, {Maiolino}, {Majewski}, {Malanushenko},
  {Malanushenko}, {Manchado}, {Mao}, {Maraston}, {Marques-Chaves}, {Masseron},
  {Masters}, {McBride}, {McDermid}, {McGrath}, {McGreer}, {Medina Pe{\~n}a}, \&
  {Melendez}}]{Blanton2017}
{Blanton}, M.~R., {Bershady}, M.~A., {Abolfathi}, B., {et~al.} 2017, \aj, 154,
  28

\bibitem[{{Boquien} {et~al.}(2019){Boquien}, {Burgarella}, {Roehlly}, {Buat},
  {Ciesla}, {Corre}, {Inoue}, \& {Salas}}]{Boquien2019}
{Boquien}, M., {Burgarella}, D., {Roehlly}, Y., {et~al.} 2019, \aap, 622, A103

\bibitem[{{Bournaud} \& {Combes}(2003)}]{Bournaud2003}
{Bournaud}, F. \& {Combes}, F. 2003, \aap, 401, 817

\bibitem[{Bradley {et~al.}(2024)Bradley, Sip{\H{o}}cz, Robitaille, Tollerud,
  Vin{\'\i}cius, Deil, Barbary, Wilson, Busko, Donath, G{\"u}nther, Cara, Lim,
  Me{\ss}linger, Burnett, Conseil, Droettboom, Bostroem, Bray, \&
  Perren}]{Bradley2024}
Bradley, L., Sip{\H{o}}cz, B., Robitaille, T., {et~al.} 2024, Zenodo, 1.12.0,
  software release

\bibitem[{{Brook} {et~al.}(2008){Brook}, {Governato}, {Quinn}, {Wadsley},
  {Brooks}, {Willman}, {Stilp}, \& {Jonsson}}]{Brook2008}
{Brook}, C.~B., {Governato}, F., {Quinn}, T., {et~al.} 2008, \apj, 689, 678

\bibitem[{{Bruzual} \& {Charlot}(2003)}]{Bruzual2003}
{Bruzual}, G. \& {Charlot}, S. 2003, \mnras, 344, 1000

\bibitem[{{Calzetti} {et~al.}(2000){Calzetti}, {Armus}, {Bohlin}, {Kinney},
  {Koornneef}, \& {Storchi-Bergmann}}]{Calzetti2000}
{Calzetti}, D., {Armus}, L., {Bohlin}, R.~C., {et~al.} 2000, \apj, 533, 682

\bibitem[{{Di Matteo} {et~al.}(2008){Di Matteo}, {Bournaud}, {Martig},
  {Combes}, {Melchior}, \& {Semelin}}]{Matteo2008}
{Di Matteo}, P., {Bournaud}, F., {Martig}, M., {et~al.} 2008, \aap, 492, 31

\bibitem[{{Dobrycheva} {et~al.}(2018){Dobrycheva}, {Vavilova}, {Melnyk}, \&
  {Elyiv}}]{Dobrycheva2018}
{Dobrycheva}, D.~V., {Vavilova}, I.~B., {Melnyk}, O.~V., \& {Elyiv}, A.~A.
  2018, Kinematics and Physics of Celestial Bodies, 34, 290

\bibitem[{{Draine} {et~al.}(2014){Draine}, {Aniano}, {Krause}, {Groves},
  {Sandstrom}, {Braun}, {Leroy}, {Klaas}, {Linz}, {Rix}, {Schinnerer},
  {Schmiedeke}, \& {Walter}}]{Draine2014}
{Draine}, B.~T., {Aniano}, G., {Krause}, O., {et~al.} 2014, \apj, 780, 172

\bibitem[{{Finkelman} {et~al.}(2012){Finkelman}, {Funes}, \&
  {Brosch}}]{Finkelman2012}
{Finkelman}, I., {Funes}, J.~G., \& {Brosch}, N. 2012, \mnras, 422, 2386

\bibitem[{{Freitas-Lemes} {et~al.}(2024){Freitas-Lemes}, {da Rocha-Poppe},
  {Fa{\'u}ndez-Abans}, {de Oliveira-Abans}, {Rodrigues}, {Tello}, \&
  {Fernandes-Martin}}]{Freitas2024}
{Freitas-Lemes}, P., {da Rocha-Poppe}, P.~C., {Fa{\'u}ndez-Abans}, M., {et~al.}
  2024, \apss, 369, 93

\bibitem[{{Freitas-Lemes} {et~al.}(2014){Freitas-Lemes}, {Rodrigues},
  {Fa{\'u}ndez-Abans}, \& {Dors}}]{Freitas2014}
{Freitas-Lemes}, P., {Rodrigues}, I., {Fa{\'u}ndez-Abans}, M., \& {Dors}, O.
  2014, in Revista Mexicana de Astronomia y Astrofisica Conference Series,
  Vol.~44, 177--177

\bibitem[{{Ghosh} {et~al.}(2020){Ghosh}, {Urry}, {Wang}, {Schawinski}, {Turp},
  \& {Powell}}]{Ghosh2020}
{Ghosh}, A., {Urry}, C.~M., {Wang}, Z., {et~al.} 2020, \apj, 895, 112

\bibitem[{Goodfellow {et~al.}(2014)Goodfellow, Pouget-Abadie, Mirza, Xu,
  Warde-Farley, Ozair, Courville, \& Bengio}]{Goodfellow2014}
Goodfellow, I., Pouget-Abadie, J., Mirza, M., {et~al.} 2014, in Advances in
  neural information processing systems, 2672--2680

\bibitem[{Guti{\'e}rrez \& Herv{\'a}s-Mart{\'i}nez(2011)}]{Gutirrez2011}
Guti{\'e}rrez, P.~A. \& Herv{\'a}s-Mart{\'i}nez, C. 2011, in Advances in
  Computational Intelligence, ed. J.~Cabestany, I.~Rojas, \& G.~Joya (Berlin,
  Heidelberg: Springer Berlin Heidelberg), 177--184

\bibitem[{{Haud}(1988)}]{Haud1988}
{Haud}, U. 1988, \aap, 198, 125

\bibitem[{{Helmi} {et~al.}(2018){Helmi}, {Babusiaux}, {Koppelman}, {Massari},
  {Veljanoski}, \& {Brown}}]{Helmi2018}
{Helmi}, A., {Babusiaux}, C., {Koppelman}, H.~H., {et~al.} 2018, \nat, 563, 85

\bibitem[{{Inoue}(2011)}]{Inoue2011}
{Inoue}, A.~K. 2011, \mnras, 415, 2920

\bibitem[{Keys(1981)}]{Keys1981}
Keys, R. 1981, IEEE Transactions on Acoustics, Speech, and Signal Processing,
  29, 1153

\bibitem[{{Khramtsov} {et~al.}(2022){Khramtsov}, {Vavilova}, {Dobrycheva},
  {Vasylenko}, {Melnyk}, {Elyiv}, {Akhmetov}, \& {Dmytrenko}}]{Khramtsov2022}
{Khramtsov}, V., {Vavilova}, I.~B., {Dobrycheva}, D.~V., {et~al.} 2022, Space
  Science and Technology, 28, 27

\bibitem[{Kingma \& Ba(2015)}]{Kingma2015}
Kingma, D. \& Ba, J. 2015, in International Conference on Learning
  Representations (ICLR), San Diega, CA, USA

\bibitem[{{Krishnakumar} \& {Bryce Kalmbach}(2022)}]{Krishnakumar2022}
{Krishnakumar}, H. \& {Bryce Kalmbach}, J. 2022, arXiv e-prints,
  arXiv:2210.11428

\bibitem[{{Lackey} {et~al.}(2023){Lackey}, {Kulkarni}, \& {Aller}}]{Lackey2023}
{Lackey}, K., {Kulkarni}, V., \& {Aller}, M. 2023, in AAS Meeting Abstracts,
  Vol.~55, AAS Meeting Abstracts, 333.02

\bibitem[{{Lackey} {et~al.}(2024){Lackey}, {Kulkarni}, \& {Aller}}]{Lackey2024}
{Lackey}, K.~E., {Kulkarni}, V.~P., \& {Aller}, M.~C. 2024, Galaxies, 12, 42

\bibitem[{{Laher} {et~al.}(2012){Laher}, {Gorjian}, {Rebull}, {Masci},
  {Fowler}, {Helou}, {Kulkarni}, \& {Law}}]{Laher2012}
{Laher}, R.~R., {Gorjian}, V., {Rebull}, L.~M., {et~al.} 2012, \pasp, 124, 737

\bibitem[{{Lauberts}(1982)}]{Lauberts1982}
{Lauberts}, A. 1982, {ESO/Uppsala survey of the ESO(B) atlas}

\bibitem[{{Leitherer} {et~al.}(2002){Leitherer}, {Li}, {Calzetti}, \&
  {Heckman}}]{Leitherer2002}
{Leitherer}, C., {Li}, I.~H., {Calzetti}, D., \& {Heckman}, T.~M. 2002, \apjs,
  140, 303

\bibitem[{Loshchilov \& Hutter(2019)}]{Loshchilov2019}
Loshchilov, I. \& Hutter, F. 2019, Decoupled Weight Decay Regularization

\bibitem[{{Macci{\`o}} {et~al.}(2006){Macci{\`o}}, {Moore}, \&
  {Stadel}}]{Maccio2006}
{Macci{\`o}}, A.~V., {Moore}, B., \& {Stadel}, J. 2006, \apjl, 636, L25

\bibitem[{{Mahdi Ali}(2021)}]{Mahdi2021}
{Mahdi Ali}, S. 2021, in Journal of Physics Conference Series, Vol. 1897,
  Journal of Physics Conference Series (IOP), 012070

\bibitem[{maintainers \& contributors(2016)}]{Torchvision2016}
maintainers, T. \& contributors. 2016, TorchVision: PyTorch's Computer Vision
  library, \url{https://github.com/pytorch/vision}

\bibitem[{{Melnyk} {et~al.}(2012){Melnyk}, {Dobrycheva}, \&
  {Vavilova}}]{Melnyk2012}
{Melnyk}, O.~V., {Dobrycheva}, D.~V., \& {Vavilova}, I.~B. 2012, Astrophysics,
  55, 293

\bibitem[{{Moiseev} {et~al.}(2011){Moiseev}, {Smirnova}, {Smirnova}, \&
  {Reshetnikov}}]{Moiseev2011}
{Moiseev}, A.~V., {Smirnova}, K.~I., {Smirnova}, A.~A., \& {Reshetnikov}, V.~P.
  2011, \mnras, 418, 244

\bibitem[{{Naidu} {et~al.}(2021){Naidu}, {Conroy}, {Bonaca}, {Zaritsky},
  {Weinberger}, {Ting}, {Caldwell}, {Tacchella}, {Han}, {Speagle}, \&
  {Cargile}}]{Naidu2021}
{Naidu}, R.~P., {Conroy}, C., {Bonaca}, A., {et~al.} 2021, \apj, 923, 92

\bibitem[{{Nilson}(1973)}]{Nilson1973}
{Nilson}, P. 1973, {Uppsala general catalogue of galaxies}

\bibitem[{{Ordenes-Brice{\~n}o} {et~al.}(2016){Ordenes-Brice{\~n}o},
  {Georgiev}, {Puzia}, {Goudfrooij}, \& {Arnaboldi}}]{Ordenes2016}
{Ordenes-Brice{\~n}o}, Y., {Georgiev}, I.~Y., {Puzia}, T.~H., {Goudfrooij}, P.,
  \& {Arnaboldi}, M. 2016, \aap, 585, A156

\bibitem[{Paszke {et~al.}(2019)Paszke, Gross, Massa, Lerer, Bradbury, Chanan,
  Killeen, Lin, Gimelshein, Antiga, Desmaison, Kopf, Yang, DeVito, Raison,
  Tejani, Chilamkurthy, Steiner, Fang, Bai, \& Chintala}]{Paszke2019}
Paszke, A., Gross, S., Massa, F., {et~al.} 2019, in Advances in Neural
  Information Processing Systems 32 (Curran Associates, Inc.), 8024--8035

\bibitem[{{Peng} {et~al.}(2002){Peng}, {Ho}, {Impey}, \& {Rix}}]{Peng2002}
{Peng}, C.~Y., {Ho}, L.~C., {Impey}, C.~D., \& {Rix}, H.-W. 2002, \aj, 124, 266

\bibitem[{Price-Whelan {et~al.}(2018)Price-Whelan, Sip{\H{o}}cz, G{\"u}nther,
  Lim, Crawford, Conseil, Shupe, Craig, Dencheva, Ginsburg,
  {et~al.}}]{Price2018}
Price-Whelan, A.~M., Sip{\H{o}}cz, B., G{\"u}nther, H., {et~al.} 2018, The
  Astronomical Journal, 156, 123

\bibitem[{{Pulatova} {et~al.}(2015){Pulatova}, {Vavilova}, {Sawangwit},
  {Babyk}, \& {Klimanov}}]{Pulatova2015}
{Pulatova}, N.~G., {Vavilova}, I.~B., {Sawangwit}, U., {Babyk}, I., \&
  {Klimanov}, S. 2015, \mnras, 447, 2209

\bibitem[{{Reshetnikov} \& {Combes}(2015)}]{Reshetnikov2015}
{Reshetnikov}, V. \& {Combes}, F. 2015, \mnras, 447, 2287

\bibitem[{{Reshetnikov} \& {Mosenkov}(2019)}]{Reshetnikov2019}
{Reshetnikov}, V.~P. \& {Mosenkov}, A.~V. 2019, \mnras, 483, 1470

\bibitem[{{Schweizer} {et~al.}(1983){Schweizer}, {Whitmore}, \&
  {Rubin}}]{Schweizer1983}
{Schweizer}, F., {Whitmore}, B.~C., \& {Rubin}, V.~C. 1983, \aj, 88, 909

\bibitem[{{Skryabina} {et~al.}(2024){Skryabina}, {Adams}, \&
  {Mosenkov}}]{Skryabina2024}
{Skryabina}, M.~N., {Adams}, K.~R., \& {Mosenkov}, A.~V. 2024, arXiv e-prints,
  arXiv:2406.13496

\bibitem[{{Smirnov} \& {Reshetnikov}(2020)}]{Smirnov2020}
{Smirnov}, D.~V. \& {Reshetnikov}, V.~P. 2020, Astronomy Letters, 46, 501

\bibitem[{{Snaith} {et~al.}(2012){Snaith}, {Gibson}, {Brook}, {Knebe},
  {Thacker}, {Quinn}, {Governato}, \& {Tissera}}]{Snaith2012}
{Snaith}, O.~N., {Gibson}, B.~K., {Brook}, C.~B., {et~al.} 2012, \mnras, 425,
  1967

\bibitem[{{Stanonik} {et~al.}(2009){Stanonik}, {Platen}, {Arag{\'o}n-Calvo},
  {van Gorkom}, {van de Weygaert}, {van der Hulst}, \&
  {Peebles}}]{Stanonik2009}
{Stanonik}, K., {Platen}, E., {Arag{\'o}n-Calvo}, M.~A., {et~al.} 2009, \apjl,
  696, L6

\bibitem[{{Stockton} \& {MacKenty}(1983)}]{Stockton1983}
{Stockton}, A. \& {MacKenty}, J.~W. 1983, \nat, 305, 678

\bibitem[{{Vavilova} {et~al.}(2020){Vavilova}, {Dobrycheva}, {Vasylenko},
  {Elyiv}, \& {Melnyk}}]{Vavilova2020}
{Vavilova}, I., {Dobrycheva}, D., {Vasylenko}, M., {Elyiv}, A., \& {Melnyk}, O.
  2020, in Knowledge Discovery in Big Data from Astronomy and Earth
  Observation, ed. P.~{{\v{S}}koda} \& F.~{Adam}, 307--323

\bibitem[{{Vavilova} {et~al.}(2021){Vavilova}, {Dobrycheva}, {Vasylenko},
  {Elyiv}, {Melnyk}, \& {Khramtsov}}]{Vavilova2021}
{Vavilova}, I.~B., {Dobrycheva}, D.~V., {Vasylenko}, M.~Y., {et~al.} 2021,
  {VizieR Online Data Catalog: SDSS galaxies morphological classification
  (Vavilova+, 2021)}

\bibitem[{{Vavilova} {et~al.}(2018){Vavilova}, {Elyiv}, \&
  {Vasylenko}}]{Vavilova2018}
{Vavilova}, I.~B., {Elyiv}, A.~A., \& {Vasylenko}, M.~Y. 2018, Radio Physics
  and Radio Astronomy, 23, 244

\bibitem[{{Vavilova} {et~al.}(2024){Vavilova}, {Fedorov}, {Dobrycheva},
  {Sergijenko}, {Vasylenko}, {Dmytrenko}, {Khramtsov}, \&
  {Kompaniiets}}]{Vavilova2024}
{Vavilova}, I.~B., {Fedorov}, P.~M., {Dobrycheva}, D.~V., {et~al.} 2024, Space
  Science and Technology, 30, 81

\bibitem[{{Vavilova} {et~al.}(2022){Vavilova}, {Khramtsov}, {Dobrycheva},
  {Vasylenko}, {Elyiv}, \& {Melnyk}}]{Vavilova2022}
{Vavilova}, I.~B., {Khramtsov}, V., {Dobrycheva}, D.~V., {et~al.} 2022, Space
  Science and Technology, 28, 03

\bibitem[{{Vavilova} {et~al.}(2023){Vavilova}, {Khramtsov}, {Dobrycheva},
  {Vasylenko}, {Elyiv}, \& {Melnyk}}]{Vavilova2023}
{Vavilova}, I.~B., {Khramtsov}, V., {Dobrycheva}, D.~V., {et~al.} 2023, VizieR
  Online Data Catalog (other), 0710, J/other/KNIT/28

\bibitem[{{Vavilova} {et~al.}(2009){Vavilova}, {Melnyk}, \&
  {Elyiv}}]{Vavilova2009}
{Vavilova}, I.~B., {Melnyk}, O.~V., \& {Elyiv}, A.~A. 2009, Astronomische
  Nachrichten, 330, 1004

\bibitem[{{Whitmore} {et~al.}(1990){Whitmore}, {Lucas}, {McElroy},
  {Steiman-Cameron}, {Sackett}, \& {Olling}}]{Whitmore1990}
{Whitmore}, B.~C., {Lucas}, R.~A., {McElroy}, D.~B., {et~al.} 1990, \aj, 100,
  1489

\bibitem[{{Xu} {et~al.}(2023){Xu}, {Hao}, {Liu}, {Lin}, {Bian}, {Hou}, {Li}, \&
  {Li}}]{Xu2023}
{Xu}, Y., {Hao}, C.~J., {Liu}, D.~J., {et~al.} 2023, \apj, 947, 54

\end{thebibliography}

\end{document}